# Randomized Self-Assembly for Exact Shapes*

David Doty

University of Western Ontario
Department of Computer Science
London, Ontario, Canada, N6A 5B7
ddoty@csd.uwo.ca

**Abstract**

Working in Winfree's abstract tile assembly model, we show that a constant-size tile assembly system can be programmed through relative tile concentrations to build an $n \times n$ square with high probability, for any sufficiently large $n$. This answers an open question of Kao and Schweller (*Randomized Self-Assembly for Approximate Shapes*, ICALP 2008), who showed how to build an *approximately* $n \times n$ square using tile concentration programming, and asked whether the approximation could be made *exact* with high probability. We show how this technique can be modified to answer another question of Kao and Schweller, by showing that a constant-size tile assembly system can be programmed through tile concentrations to assemble arbitrary finite *scaled shapes*, which are shapes modified by replacing each point with a $c \times c$ block of points, for some integer $c$. Furthermore, we exhibit a smooth tradeoff between specifying bits of $n$ via tile concentrations versus specifying them via hard-coded tile types, which allows tile concentration programming to be employed for specifying a fraction of the bits of "input" to a tile assembly system, under the constraint that concentrations can only be specified to a limited precision. Finally, to account for some unrealistic aspects of the tile concentration programming model, we show how to modify the construction to use only concentrations that are arbitrarily close to uniform.

## 1 Introduction

Self-assembly is a term used to describe systems in which a small number of simple components, each following local rules governing their interaction with each other, automatically assemble to form a target structure. Winfree [27] introduced the abstract Tile Assembly Model (aTAM) – based on a constructive version of Wang tiling [25, 26] – as a simplified mathematical model of Seeman's work [20] in utilizing DNA to physically implement self-assembly at the molecular level. In the aTAM, the fundamental components are un-rotatable, but translatable square "tile types" whose sides are labeled with glue "labels" and "strengths." Two tiles placed next to each other *interact* if the glue labels on their abutting sides match, and a tile *binds* to an assembly if the total strength on all of its interacting sides exceeds the ambient "temperature," equal to 2 in this paper. The model is detailed more formally in Section 2.

*A preliminary version of this article appeared as [9].

Winfree [27] demonstrated the computational universality of the aTAM by showing how to simulate an arbitrary cellular automaton with a tile assembly system. Building on these connections to computability, Rothemund and Winfree [17] investigated the minimum number of tile types needed to uniquely assemble an $n \times n$ square. Utilizing the theory of Kolmogorov complexity, they show that for any algorithmically random $n$, $\Omega\left(\frac{\log n}{\log\log n}\right)$ tile types are required to uniquely assemble an $n \times n$ square, and Adleman, Cheng, Goel, and Huang [1] exhibit a construction showing that this lower bound is asymptotically tight.

Real-life implementations of the aTAM involve (at the present time) creating tile types out of DNA double-crossover molecules [18], copies of which can be created at an exponential rate using the polymerase chain reaction (PCR) [19]. PCR technology has advanced to the point where it is automated by machines, meaning that *copies* of tiles are easy to supply, whereas the number of distinct tile *types* is a precious resource, costing much more lab time to create. Therefore, effort has been put towards developing methods of "programming" tile sets through methods other than hard-coding the desired behavior into the tile types. Such methods include *temperature programming* [10, 24], which involves changing the ambient temperature through the assembly process in order to alter which bonds are possible to break or create, and *staged assembly* [8], which involves preparing different assemblies in different test tubes, which are then mixed after reaching a terminal state. Each of these models allows a *single* tile set to be reused for assembling different structures by programming different environmental conditions that affect the behavior of the tiles and therefore serve as an "input" to be processed by the tile set.

The "input specification model" used in this paper is known as *tile concentration programming.* If the tile assembly system is nondeterministic – if intermediate assemblies exist in which more than one tile type is capable of binding to the same position – and if the solution is well-mixed, then the relative concentrations of these tile types determine the probability that each tile type will be the one to bind. Tile concentrations affect the expected time before an assembly is completed (such a model is considered in [1] and [2], for instance), but we ignore such running time considerations in the present paper. We instead focus on using the biased randomness of tile concentrations to guide a probabilistic shape-building algorithm, subject a certain kind of "geometric *space* bound"; namely, that the algorithm must be executed within the confines of the shape being assembled. This restriction follows from the monotone nature of the aTAM: once a tile attaches to an assembly, it never detaches.

We now describe related work. Chandran, Gopalkrishnan, and Reif [5] show that a one-dimensional line of expected length $n$ can be assembled using $\Theta(\log n)$ tile types, subject to the restriction that all tile concentrations are equal. Furthermore, they show that this bound is tight for *all* $n$. Note that this is not tile concentration programming since the concentrations are forced to be equal. Nonetheless, they use the inherent randomness of binding competition to strictly improve the assembly capabilities of the aTAM; a simple pigeonhole argument shows that $n$ unique tile types are required to construct a line of length $n$ in the deterministic aTAM model. Two previous papers [2, 11] deal directly with the tile concentration programming model. Becker, Rapaport, and Rémila [2] show that there is a *single* tile assembly system $\mathcal{T}$ such that, for all $n \in \mathbb{N}$, setting the tile concentrations appropriately causes $\mathcal{T}$ to assemble an $n' \times n'$ square, such that $n'$ has expected value $n$. However, $n'$ will have a large deviation from $n$ with non-negligible probability. Kao and Schweller [11] improve this result by constructing, for each $\delta, \epsilon > 0$, a tile assembly system $\mathcal{T}$ such that setting the tile concentrations appropriately causes $\mathcal{T}$ to assemble an $n' \times n'$ square, where $(1-\epsilon)n \leq n' \leq (1+\epsilon)n$ with probability at least $1-\delta$, for sufficiently large $n \in \mathbb{Z}^+$ (depending on

$\delta$ and $\epsilon$).

Kao and Schweller asked whether a constant-sized tile assembly system could be constructed that, through tile concentration programming, would assemble a square of dimensions *exactly* $n \times n$, with high probability. We answer this question affirmatively, showing that, for each $\delta > 0$, there is a tile assembly system $\mathcal{T}$ such that, for sufficiently large $n \in \mathbb{Z}^+$, there is an assignment of tile concentrations to $\mathcal{T}$ such that $\mathcal{T}$ assembles an $n \times n$ square with probability at least $1 - \delta$. Therefore, with a constant number of tile types, any size square can be created entirely through the programming of tile concentrations. The primary technique is a tile set that, through appropriate tile concentration programming, forms a thin structure of height $O(\log n)$ and length $O(n^{2/3})$ (and for arbitrarily small $\epsilon > 0$, the length can be made $O(n^\epsilon)$) and encodes the value of $n$ in binary. This binary string could be used to assemble useful structures other than squares, such as rectangles and other supersets of the sampling structure that are "easily encoded" in a binary string of length $O(\log n)$.

Kao and Schweller also asked whether arbitrary finite connected shapes, possibly scaled by factor $c \in \mathbb{N}$ (depending on the shape) by replacing each point in the shape with a $c \times c$ block of points, could be assembled from a constant tile set through concentration programming. Our construction answering the first question computes the binary expansion of $n$ with high probability in a self-assembled rectangle of height $O(\log n)$ and width $O(n^{2/3})$. By assembling this structure within the "seed block" of the construction of [23], our construction can easily be combined with that of [23] to answer this question affirmatively as well, by replacing the number $n$ with a program that outputs a list of points in the shape, and using this as the "seed block" of the construction of [23].

Since it may be infeasible to specify tile concentrations with unlimited precision, we show how to generalize our construction to allow a smooth tradeoff between specifying the number $n$ through tile concentrations versus hard-coded tile types. Since $\log n$ bits are required to specify $n$ for almost all values of $n$, we show that for arbitrary $g$, it is possible to specify "about" $g$ of the bits through tile concentrations and the remaining "about" $\log n - g$ bits through the hard-coding of tile types; i.e., using a tile set that can be described with about $(\log n) - g + o(\log n)$ bits. The actual bound is complicated and is stated in Theorem 5.3.

Finally, there are some unrealistic aspects of the concentration programming model, in addition to the assumption that concentrations can be specified to unlimited precision. Chiefly, the aTAM is itself an kinetically implausible model, but Winfree showed that the behavior of the aTAM can be approximated to arbitrary accuracy by the more realistic *kinetic Tile Assembly Model* (kTAM) [27]. One of the assumptions Winfree employs to achieve this approximation is that all tile types have equal concentration, a condition clearly violated by our intentional setting of concentrations to be unequal. We will argue that our particular construction avoids the potential pitfalls of the concentration programming model, but leave open the task of defining a concentration programming model that is inherently immune to these pitfalls. We also show how to alter the construction to use only concentrations that are arbitrarily close to uniform, as a potential fix for the kinetic problems.

This paper is organized as follows. Section 2 provides background definitions and notation for the abstract TAM and tile concentration programming. Section 3 specifies and proves the correctness of the main construction, a tile set that can be used to assemble precisely-sized squares through concentration programming. This results of Section 3 first appeared in [9]. Section 4 specifies the construction of a tile set that assembles scaled versions of arbitrary finite shapes through concentration programming. This scaled shapes construction was announced in [9] but not demonstrated. Section 5 discusses a relaxation of the model that allows a greater than constant number of tile types, while using fewer bits to specify the concentrations of each tile type, and shows how to achieve a smooth tradeoff between the resources of "number of tile types" and "bits of precision of concentrations", to assemble squares, while using an asymptotically optimal number of total bits of precision (i.e., the bits of precision of concentrations, plus the bits needed to describe the tile types, are $O(\log n)$ for an $n \times n$ square). Section 6 discusses some unrealistic aspects of the concentration programming model, and argues that the constructions of this paper are resistant to the problems caused by those unrealistic assumptions, or can be fixed to alleviate these problems. Section 7 concludes the paper, discusses practical limitations of the construction and potential improvements and states open questions.

## 2 The Tile Assembly Model and Tile Concentration Programming

We give a brief sketch of the Tile Assembly Model. More details and discussion may be found in [12, 16, 17, 27]. Our notation is that of [12], which provides a more detailed and self-contained introduction to the Tile Assembly Model for the reader unfamiliar with the model.

All logarithms in this paper are base 2. We work in the 2-dimensional discrete space $\mathbb{Z}^2$. Define the set $U_2 = \{(0,1),(1,0),(0,-1),(-1,0)\}$ to be the set of all *unit vectors*, i.e., vectors of length 1 in $\mathbb{Z}^2$. We write $[X]^2$ for the set of all 2-element subsets of a set $X$. All *graphs* in this paper are undirected graphs, i.e., ordered pairs $G = (V, E)$, where $V$ is the set of *vertices* and $E \subseteq [V]^2$ is the set of *edges*.

Intuitively, a tile type $t$ is a unit square that can be translated, but not rotated, having a well-defined "side $\vec{u}$" for each $\vec{u} \in U_2$. Each side $\vec{u}$ of $t$ has a "glue" with "label" $\text{label}_t(\vec{u})$ – a string over some fixed alphabet $\Sigma$ – and "strength" $\text{str}_t(\vec{u})$ – a nonnegative integer – specified by its type $t$. Two tiles $t$ and $t'$ that are placed at the points $\vec{a}$ and $\vec{a} + \vec{u}$ respectively, *bind* with *strength* $\text{str}_t(\vec{u})$ if and only if $(\text{label}_t(\vec{u}), \text{str}_t(\vec{u})) = (\text{label}_{t'}(-\vec{u}), \text{str}_{t'}(-\vec{u}))$. In our figures, we follow the convention of representing strength-0 bonds with dashed lines, strength-1 bonds with single lines, and strength-2 bonds with double lines.

Given a set $T$ of tile types, an *assembly* is a partial function $\alpha : \mathbb{Z}^2 \dashrightarrow T$, with points $\vec{x} \in \mathbb{Z}^2$ at which $\alpha(\vec{x})$ is undefined interpreted to be empty space, so that dom $\alpha$ is the set of points with tiles. We write $|\alpha|$ to denote $|\text{dom } \alpha|$, and we say $\alpha$ is *finite* if $|\alpha|$ is finite. For assemblies $\alpha$ and $\beta$, we say that $\alpha$ is a *subassembly* of $\beta$, and write $\alpha \sqsubseteq \beta$, if dom $\alpha \subseteq$ dom $\beta$ and $\alpha(\vec{x}) = \beta(\vec{x})$ for all $\vec{x} \in$ dom $\alpha$. $\beta$ is a *single-tile extension* of $\alpha$ if $\alpha \sqsubseteq \beta$ and dom $\beta \setminus$ dom $\alpha$ is a singleton set. In this case, we write $\beta = \alpha + (\vec{m} \mapsto t)$, where $\{\vec{m}\} = \text{dom } \beta \setminus \text{dom } \alpha$ and $t = \beta(\vec{m})$.

A *grid graph* is a graph $G = (V, E)$ in which $V \subseteq \mathbb{Z}^2$ and every edge $\{\vec{a}, \vec{b}\} \in E$ has the property that $\vec{a} - \vec{b} \in U_2$. The *binding graph of* an assembly $\alpha$ is the grid graph $G_\alpha = (V, E)$, where $V =$ dom $\alpha$, and $\{\vec{m}, \vec{n}\} \in E$ if and only if (1) $\vec{m} - \vec{n} \in U_2$, and (2) $\alpha(\vec{m})$ and $\alpha(\vec{n})$ bind with positive strength. An assembly is *$\tau$-stable*, where $\tau \in \mathbb{N}$, if it cannot be broken up into smaller assemblies without breaking bonds of total strength at least $\tau$; i.e., if every cut of $G_\alpha$ has weight at least $\tau$, where the weight of an edge is the strength of the glue it represents. In contrast to the model of Wang tiling, the nonnegativity of the strength function implies that glue mismatches between adjacent tiles do not prevent a tile from binding to an assembly, so long as sufficient binding strength is received from the sides of the tile at which the glues match. The *frontier* of an assembly $\alpha$ is $\partial\alpha = \bigcup_{t \in T} \{ \ \vec{m} \ \mid \vec{m} \notin \text{dom } \alpha \text{ and } \alpha + (\vec{m} \mapsto t) \text{ is } \tau\text{-stable} \ \}$, the set of locations at which a single tile could be stably added to $\alpha$.

Self-assembly begins with a *seed assembly* $\sigma$ (typically assumed to be finite and $\tau$-stable) and proceeds asynchronously and nondeterministically,[1] with tiles adsorbing one at a time to the existing assembly in any manner that preserves stability at all times, formally modeled as follows.

A *tile assembly system* (*TAS*) is an ordered triple $\mathcal{T} = (T, \sigma, \tau)$, where $T$ is a finite set of tile types, $\sigma : \mathbb{Z}^2 \dashrightarrow T$ is the finite, $\tau$-stable *seed assembly*, and $\tau \in \mathbb{N}$ is the *temperature*, equal to 2 in this paper.[2] $\mathcal{T}$ is *singly-seed* if $|\text{dom } \sigma| = 1$. An *assembly sequence* of a TAS $\mathcal{T} = (T, \sigma, 2)$ is a (finite or countably infinite) sequence $\vec{\alpha} = (\alpha_i \mid 0 \leq i < k)$ (with $k \in \mathbb{N} \cup \{\infty\}$) of assemblies in which $\alpha_0 = \sigma$, each $\alpha_{i+1}$ is a single-tile extension of $\alpha_i$, and each $\alpha_i$ is $\tau$-stable. The *result* of $\vec{\alpha}$ is the unique assembly $\text{res}(\vec{\alpha})$ such that $\text{dom res}(\vec{\alpha}) = \bigcup_{i=0}^{k-1} \text{dom } \alpha_i$ and, for all $0 \leq i < k$, $\alpha_i \sqsubseteq \text{res}(\vec{\alpha})$. In the case that $k$ is finite, it is routine to verify that $\text{res}(\vec{\alpha}) = \alpha_{k-1}$. We write $\mathcal{A}[\mathcal{T}]$ to denote the set of all results of assembly sequences of $\mathcal{T}$ starting with the seed assembly, known as the *producible assemblies* of $\mathcal{T}$. An assembly $\alpha$ is *terminal* if no tile can be stably added to it; i.e., if $\partial\alpha = \varnothing$. If $\alpha$ is producible and terminal, we write $\alpha \in \mathcal{A}_\Box[\mathcal{T}]$. An assembly sequence $\vec{\alpha} = (\alpha_i \mid 0 \leq i < k)$ is *fair* if for all $i$ and all $\vec{m} \in \partial\alpha_i$, there exists $j$ such that $\alpha_j(\vec{m})$ is defined; i.e., no frontier location is "starved". It is routine to verify that $\vec{\alpha}$ is fair if and only if $\text{res}(\vec{\alpha})$ is terminal.

A *tile concentration assignment* on $\mathcal{T}$ is a function $\rho : T \to [0, \infty)$.[3] If $\rho(t)$ is not specified explicitly for some $t \in T$, then $\rho(t) = 1$. $\rho$ induces a probability measure $P_\rho : \mathcal{A}_\Box[\mathcal{T}] \to [0, 1]$ in the following way. Let $\alpha \in \mathcal{A}_\Box[\mathcal{T}]$ be a producible, terminal assembly. Let $A(\alpha)$ be the set of all assembly sequences $\vec{\alpha} = (\alpha_i \mid 0 \leq i < k)$ such that $\text{res}(\vec{\alpha}) = \alpha$. Write $T_{\alpha_i}(\vec{m}) = \{\ t \in T \mid \alpha_i + (\vec{m} \mapsto t) \text{ is } \tau\text{-stable}\ \}$ for the set of tile types $t$ that are stably attachable at position $\vec{m} \in \partial\alpha_i$. Let $f_{\alpha_i}(\vec{m}) = \sum\limits_{t \in T_{\alpha_i}(\vec{m})} \rho(t)$. Define the *frontier selection probability*

$$p_{\alpha_i}(\vec{m}) = \frac{f_{\alpha_i}(\vec{m})}{\sum\limits_{\vec{n} \in \partial\alpha_i} f_{\alpha_i}(\vec{n})}.$$

This quantity is the probability that $\vec{m}$ is the position of next attachment to $\alpha_i$. Let $t \in T_{\alpha_i}(\vec{m})$, and define the *tile selection probability*

$$p_{\alpha_i}(t|\vec{m}) = \frac{\rho(t)}{f_{\alpha_i}(\vec{m})}.$$

This quantity is the *conditional* probability that $t$ attaches to position $\vec{m}$ of $\alpha_i$, *given* that $\vec{m} \in \partial\alpha_i$ is the frontier location that is tiled at stage $i$ of assembly. Define $\vec{m}_{\vec{\alpha},i} \in \partial\alpha_i$ to be the frontier location that is tiled in $\alpha_i$ to create $\alpha_{i+1}$, and let $t_{\vec{\alpha},i} = \alpha_{i+1}(\vec{m})$ be the tile type placed there. We

[1] There are multiple senses in which a tile system can be nondeterministic. One sense is that the location of attachment, if there is more than one candidate, is selected nondeterministically. Such systems may still be deterministic in the sense that they will lead to a unique final assembly. We employ a stronger version of nondeterminism in which the tile capable of binding to a *single* position of an assembly is not fixed; the randomized algorithm we implement relies on this choice being made according to the tile concentrations.

[2] A tile set can be "programmed" with different inputs through selection of an appropriate seed assembly. In this paper, we wish to model the situation in which, once work has been done once to create a single tile set, the tile set can be programmed *entirely* through adjustment of tile concentrations. Hence, our result is stated in terms of the existence of a tile assembly *system*, with a fixed seed assembly (in fact, a single seed tile), that can be used to construct squares of any size, solely by adjusting the tile concentrations.

[3] Note in particular that we do not require $\rho$ to be a probability measure on $T$.

define the probability measure $P_\rho : \mathcal{A}_\Box[\mathcal{T}] \to [0,1]$ as

$$P_\rho(\alpha) = \sum_{\vec{\alpha}=(\alpha_i|0\le i<k)\in A(\alpha)} \prod_{i=0}^{k-2} p_{\alpha_i}(\vec{m}_{\vec{\alpha},i}) p_{\alpha_i}(t_{\vec{\alpha},i}|\vec{m}_{\vec{\alpha},i}).$$

By the identity $\Pr(A \text{ and } B) = \Pr(A)\Pr(B|A)$, the quantity $p_{\alpha_i}(\vec{m}_{\vec{\alpha},i}) p_{\alpha_i}(t_{\vec{\alpha},i}|\vec{m}_{\vec{\alpha},i})$ is the probability that $\vec{m}_{\vec{\alpha},i}$ is the next location of attachment and that $t_{\vec{\alpha},i}$ is the tile type to be placed there. For an event $E$, write $\Pr_{\alpha \xleftarrow{P_\rho} \mathcal{A}_\Box[\mathcal{T}]}[E]$ to denote the probability that $E$ happens when $\alpha$ is sampled according to distribution $P_\rho$.

For $A, B \subseteq \mathbb{Z}^2$ and $\vec{u} \in \mathbb{Z}^2$, we write $A + \vec{u}$ to denote the set $\{\ \vec{v} + \vec{u} \mid \vec{v} \in A\ \}$, and we write $A \simeq B$ if there exists $\vec{u}$ such that $A + \vec{u} = B$; i.e., if $A$ is a translation of $B$. For $p \in [0,1]$ and $X \subseteq \mathbb{Z}^2$, we say $X$ *strictly self-assembles in* $\mathcal{T}(\rho)$ *with probability at least* $p$ if $\Pr_{\alpha \xleftarrow{P_\rho} \mathcal{A}_\Box[\mathcal{T}]}[\text{dom}\ \alpha \simeq X] \ge p$. That is, $\mathcal{T}$ self-assembles into the same shape as $X$ with probability at least $p$. Note that two different assemblies may have the same shape though they might assign different tile types to the same position.

The definition of $P_\rho$ takes into account not only the tile selection probability, the effect of tile concentrations on which tile type is selected when more than one compete to bind to a single frontier location, but also the frontier selection probability, which of multiple frontier locations is selected. However, all constructions in this paper are correct so long as the assembly sequence is fair. By the following observation, the assembly sequence can be assumed fair so long as all concentrations are strictly positive, implying that we need not consider the frontier selection probability when arguing the correctness of the constructions. Obviously, there are finite assembly sequences $\vec{\alpha}$ occurring with positive probability such that $\text{res}(\vec{\alpha})$ is not terminal. However, we want to establish that as long as growth is allowed to continue whenever the assembly is nonterminal, the probability of the assembly sequence being fair is 1. Therefore the observation is stated only for infinite assembly sequences.

**Observation 2.1.** *Let* $\mathcal{T} = (T, \sigma, \tau)$ *be a TAS, let* $\rho : T \to (0, \infty)$ *be a strictly positive tile concentration assignment, and let* $\vec{\alpha}$ *be an infinite assembly sequence resulting from the assembly of* $\mathcal{T}$ *according to* $\rho$ *as described above. Then* $\Pr[\vec{\alpha} \text{ is fair}] = 1$.

*Proof.* Let $\vec{\alpha} = (\alpha_i \mid 0 \le i < \infty)$, where $\alpha_0 = \sigma$, and consider assembly $\alpha_i$ for some $i$. Since each tile addition increases the size of the frontier by at most 3, $|\partial\alpha_i| \le 3i + |\partial\sigma|$. Define $f_{\min} = \min_{t\in T} \rho(t)$, $f_{\max} = \sum_{t\in T} \rho(t)$, and $f_{\text{ratio}} = f_{\min}/f_{\max}$. Let $0 \le p < k$ be some stage where $\alpha_p$ is not terminal, and let $\vec{m} \in \partial\alpha_p$. It suffices to show that $\Pr[\vec{m} \text{ is never tiled}] = 0$. Note that $f_{\min} \le f_{\alpha_i}(\vec{m}) \le f_{\max}$ for all $i$. Then we have

$$p_{\alpha_i}(\vec{m}) = \frac{f_{\alpha_i}(\vec{m})}{\sum\limits_{\vec{n}\in\partial\alpha_i} f_{\alpha_i}(\vec{n})} \ge \frac{f_{\min}}{\sum\limits_{\vec{n}\in\partial\alpha_i} f_{\max}} = \frac{f_{\text{ratio}}}{|\partial\alpha_i|} \ge \frac{f_{\text{ratio}}}{3i + |\partial\sigma|}.$$

Then by the inequality $1 + x \le e^x$ for all $x \in \mathbb{R}$,

$$\begin{aligned}
\Pr[\vec{m} \text{ is never tiled}] &= \prod_{i=p}^{\infty}(1 - p_{\alpha_i}(\vec{m})) \quad \le \quad \prod_{i=p}^{\infty} e^{-p_{\alpha_i}(\vec{m})} \\
&\le \prod_{i=p}^{\infty} e^{-\frac{f_{\text{ratio}}}{3i+|\partial\sigma|}} \quad = \quad e^{-f_{\text{ratio}} \sum_{i=p}^{\infty} \frac{1}{3i+|\partial\sigma|}},
\end{aligned}$$

which is equal to 0 since the sum is a divergent general harmonic series. □

# 3 Constructing a Square using $O(1)$ Tile Types by Tile Concentration Programming

This section is devoted to proving the following theorem, which is the main result of this paper.

For all $\delta > 0$ and $n \in \mathbb{N}$, define $r_\delta = \left\lceil \frac{\log \frac{\delta}{8}}{\log 0.9421} \right\rceil$, $c_\delta = 2 + \left\lceil \log \frac{\log \frac{\delta}{8}}{\log 0.717} \right\rceil$, and $k_n = \left\lceil \frac{\lfloor \log n \rfloor + 1}{3} \right\rceil$, and define $b(n, \delta) = \max\left\{ r_\delta, 2^{2k_n + c_\delta} \right\} + c_\delta + 3k_n$.

**Theorem 3.1.** *For all $\delta > 0$, there is a tile assembly system $\mathcal{T}_\delta = (T, \sigma, 2)$ such that, for all integers $n \geq b(n, \delta)$, there is a tile concentration assignment $\rho_n : T \to [0, \infty)$ such that the set $\{\ (x, y) \mid x, y \in \{1, \ldots, n\}\ \}$ strictly self-assembles in $\mathcal{T}_\delta(\rho_n)$ with probability at least $1 - \delta$.*

Note that for any fixed $\delta > 0$, $b(n, \delta) = O(n^{2/3})$ (where the constant in the $O()$ depends on $\delta$), whence $n \geq b(n, \delta)$ for all sufficiently large $n$.

## 3.1 Intuitive Idea of the Construction

Kao and Schweller introduced a basic primitive in [11] (refining a lower-precision technique described in [2]), called a *sampling line*. The sampling line allows tile concentrations to encode a natural number whose binary representation can be probably approximately reproduced. Kao and Schweller utilize the sampling line to encode $n \in \mathbb{N}$ by an approximation $n' \in \mathbb{N}$ such that $(1 - \epsilon)n \leq n' \leq (1 + \epsilon)n$ with probability at least $1 - \delta$.

The idea of our construction is as follows. We will "approximate" only numbers $m$ small enough that the sampling line approximation has sufficient space to be an *exact* computation of $m$ with high probability. The construction of Kao and Schweller can be thought of as estimating $n$ by, in a sense, probabilistically counting to $n$ using independent Bernoulli trials with appropriately fixed success probability; i.e., the probabilities are used to estimate an approximate *unary* encoding of $n$, which is converted to binary by a counter. Representing $n$ in unary, of course, takes space $n$, and recovering it probabilistically from tiles subject to randomization requires using much more than space $n$ to overcome the error introduced by randomization. Kao and Schweller use an ingenious technique to spread this estimation out into the center of the $n \times n$ square being built, affording $O(n^2)$ space to approximate $n$ closely. However, that construction lacks the space to compute $n$ exactly, which requires much more than $n^2$ Bernoulli trials – applying the standard Chernoff bound to the Kao-Schweller sampling line achieves an upper bound of $O(n^5)$ trials – to achieve a sufficiently small estimation error. Hence, attempting to use a sampling line directly to compute $n$ would result in a line containing many more tiles than the $n^2$ tiles that compose an $n \times n$ square, and no amount of twisting the line will cause it to fit inside the boundaries of the square.

We split $n$'s binary expansion $b(n) = b_1 b_2 \ldots b_{\lfloor \log n \rfloor + 1} \in \{0, 1\}^*$ into three subsequences $b_1 b_4 b_7 \ldots$, $b_2 b_5 b_8 \ldots$, and $b_3 b_6 b_9 \ldots$, each of length about $\frac{1}{3} \log n$, and interpret these binary strings as natural numbers $m_1, m_2, m_3 \leq n^{1/3}$ to be estimated. The problem of estimating $n$ is reduced to that of estimating these three numbers. At the same time, we introduce a new sampling line technique that can exactly estimate a number $m$ with high probability using only $O(m^2)$ trials.[4] Since

[4]As opposed to the $O(m^5)$ trials that would be required by the Kao-Schweller sampling line. It is possible to use Kao and Schweller's original sampling line to estimate seven numbers – $\lfloor \log n \rfloor + 1$ (the length of the binary expansion of $n$), and the six numbers $m_1$ - $m_6$ encoded by length-$\left\lceil \frac{\lfloor \log n \rfloor + 1}{6} \right\rceil$ substrings of $n$'s binary expansion, each small enough that $m_i^5 = o(n)$ – and to use these numbers to reconstruct $n$ and from that, build an $n \times n$ square. A

$m_1, m_2, m_3 \leq n^{1/3}$, estimating $m_1, m_2$, and $m_3$ will require $O(n^{2/3})$ trials, which fits within the width of an $n \times n$ square for sufficiently large $n$.

Intuitively, the reason that estimating $m_1$, $m_2$, and $m_3$ creates an improvement over estimating $n$ directly is that the space needed for the unary encodings of numbers whose *binary* length is one-third that of $n$'s does not scale linearly with that length; the *unary* encoding of these numbers scales with $n^{1/3}$, not $n/3$, whence a quadratic increase in the space needed for probabilistic recovery remains sufficiently small ($O(n^{2/3})$) that three such decodings easily fit into space $n$.

### 3.2 Probabilistic Decoding of a Natural Number using a Sampling Line

In this section, we describe how to exactly compute a positive integer $m$ probabilistically from tile concentrations that are appropriately programmed to represent $m$. In our final construction, the sampling line will estimate not one but three integers $m_1$, $m_2$, and $m_3$, as described in Section 3.1, by embedding additional bits into the tiles. However, for the sake of clarity, in this section, we describe how to estimate a single positive integer $m$, and then describe in Section 3.2.2 how to modify the construction and set the probabilities to allow three numbers to be estimated simultaneously on a single sampling line.

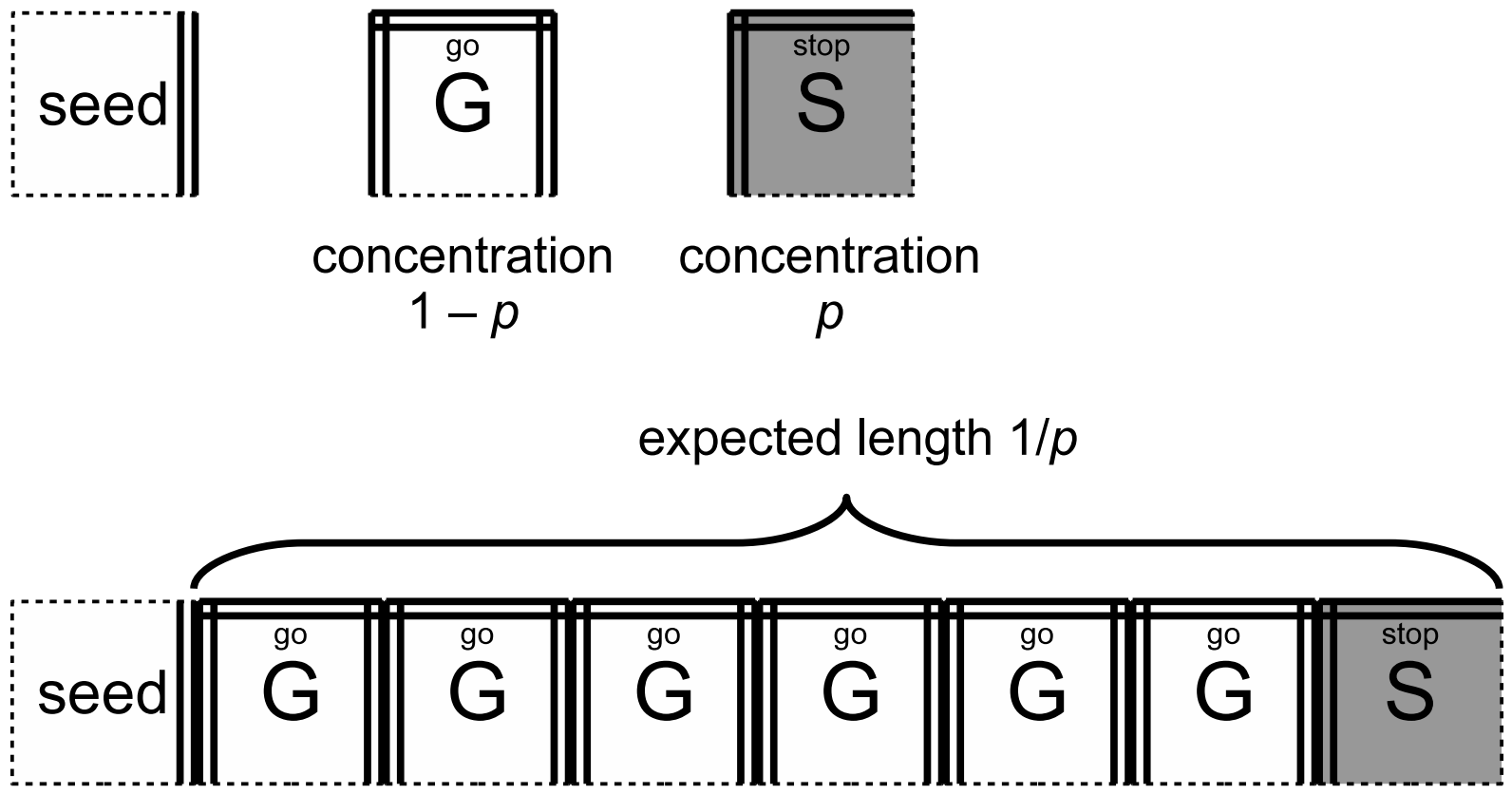

Figure 1: The portion of the basic Kao-Schweller sampling line that controls its length. Two tiles compete nondeterministically to bind to the right of the line, one of which stops the growth, while the other continues, giving the length of the line a geometric distribution.

The basic length-controlling portion of the Kao-Schweller sampling line is shown in Figure 1.[5] A horizontal row of tiles forms to the right of the seed. Two tiles, $G$ ("go") and $S$ ("stop") nondeterministically connect to the right end of the line; $G$ continues the growth, while $S$ stops the growth. If $S$ has concentration $p \in [0,1]$ and $G$ has concentration $1-p$, then the length $L$ of the line is a geometric random variable with expected value $1/p$. By setting $p$ appropriately, $\mathrm{E}[L]$ straightforward and tedious analysis of the constants involved reveals that such a technique can be used to construct $n \times n$ squares for $n \geq 10^{18}$. We achieve much more feasible bounds on $n$ ($\approx 10^7$ for $\delta = 0.01$) using the techniques introduced in this paper, and indeed, better bounds than those required by Kao and Schweller to approximate $n$, whose construction achieves, for instance, a $(0.01, 0.01)$-approximation only for $n \geq 10^{13}$, according to their analysis.

[5]Our description of the Kao-Schweller sampling line is incomplete, as discussed in the next paragraph.

can be controlled, but not precisely, since a geometric random variable may have a deviation from the expected value that is too large for our purposes.

Kao and Schweller allow a third tile type to bind within the sampling line, which does the actual sampling for computing a natural number, but our construction splits this sampling into a separate set of tiles that forms above the line. The sampling portion is discussed in Section 3.2.2. For the present time, we restrict our discussion to controlling the length of the line.

### 3.2.1 More Precisely Controlling the Sampling Line Length

Our goal is to control $L$, the length of the sampling line, such that, by setting tile concentrations appropriately, we may ensure that $L$ lies between $2^{a-1}$ and $2^a$ with high probability, for an $a \in \mathbb{Z}^+$ of our choosing (which will be influenced by the number $n$ we are estimating). That is, we may ensure that the number of bits required to represent $L$ is computed precisely, even if the exact value of $L$ varies widely within the interval $[2^{a-1}, 2^a)$. We then attach a counter – a group of tiles that measures the length of the line by counting in binary – to the north of the line that measures $L$ until the final stopping tile. The stop signal is not intended to stop the counter immediately, but rather to signal that the counter should continue until it reaches the next power of 2 – i.e., the next time a new most significant bit is required – and then stop. Hence, we may choose an arbitrary power of 2 and set tile concentrations to ensure that the counter counts to that value and then stops.

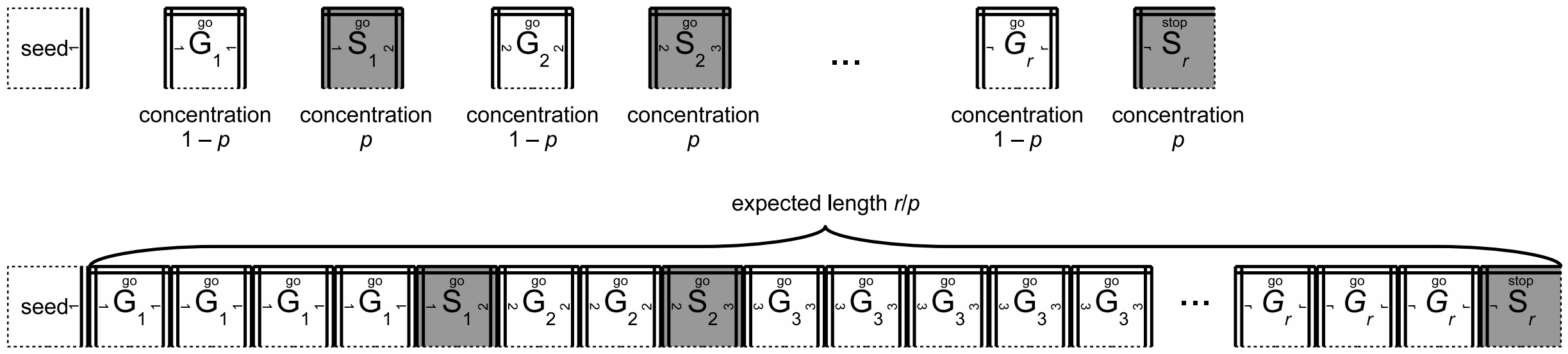

Figure 2: The portion of the sampling line of our construction that controls its length. $r$ stages each have expected length $1/p$, making the expected total length $r/p$, but more tightly concentrated about that expected length than in the case of one stage.

To increase the precision with which we control $L$, we use not one but many stages of "go" and "stop" tiles, $G_1, S_1, G_2, S_2, \ldots, G_r, S_r$. The construction is shown in Figure 2. $G_i$ and $S_i$ each compete to bind to the right of $S_{i-1}$ and $G_i$. $S_i$ signals a transition to the next stage $i+1$, with $S_r$ stopping the growth of the line after $r$ stages. Therefore, the sequence of tiles to the right of the seed is a string described by the regular expression $G_1^* S_1 G_2^* S_2 \ldots G_r^* S_r$. Each $S_i$ has concentration $p$, and the remaining $G_i$ tiles each have concentration $1-p$. The length $L$ of the line is a *negative binomial* random variable[6] with parameters $r, p$ (see [14]) with expected value $r/p$ by linearity of expectation; i.e., its length is the number of Bernoulli trials required before exactly $r$ successes, provided each Bernoulli trial has success probability $p$.

[6]The term *negative* is misleading; a negative binomial random variable is better described (informally) as the *inverse* of a binomial random variable, if one thinks of a binomial random variable as being like a function that maps a number of Bernoulli trials to a number of successes. A negative binomial random variable maps a number of successes to the number of trials necessary to achieve that number of successes.

Let $N, R \in \mathbb{N}$ and $p \in [0,1]$. A binomial random variable $\mathcal{B}(N,p)$ (the number of successes after $N$ Bernoulli trials, each having success probability $p$) is related to a negative binomial random variable $\mathcal{N}(R,p)$ (the number of trials before exactly $R$ successes) by the relationships

$$\Pr[\mathcal{N}(R,p) < N] = \Pr[\mathcal{B}(N,p) > R] \text{ and} \tag{3.1}$$

$$\Pr[\mathcal{N}(R,p) > N] = \Pr[\mathcal{B}(N,p) < R]. \tag{3.2}$$

Thus, Chernoff bounds that provide tail bounds for binomial distributions can be applied to negative binomial distributions via (3.1) and (3.2).

To cause $L$ to fall in the interval $[2^{a-1}, 2^a)$, we must set its expected length $\overline{L}$ (by setting $p = r/\overline{L}$) to be such that the $r^{\text{th}}$ success occurs when the line has length in the interval $[2^{a-1}, 2^a)$. Note that $pN$ is the expected number of successes in the first $N$ tiles of the line; i.e., it is the expected number of successes in exactly $N$ Bernoulli trials.

We define $\epsilon$ and $\epsilon'$ so that $\overline{L} = (1+\epsilon)2^{a-1} = (1-\epsilon')2^a$ and the two error probabilities derived below are approximately equal; $\epsilon \approx 0.442695$ and $\epsilon' \approx 0.2786525$ suffice. The event that $L < 2^{a-1}$ is equivalent to the event that $2^{a-1}$ Bernoulli trials are conducted (with expected number of successes $p2^{a-1}$) with at least $r$ successes. By (3.1) and the Chernoff bound [14, Theorem 4.4, part 1],

$$\begin{aligned}\Pr\left[L < 2^{a-1}\right] &= \Pr\left[r > (1+\epsilon)p2^{a-1}\right] &&\leq \left(\tfrac{e^{\epsilon}}{(1+\epsilon)^{1+\epsilon}}\right)^{p2^{a-1}} &&= \left(\tfrac{e^{\epsilon}}{(1+\epsilon)^{1+\epsilon}}\right)^{r2^{a-1}/\overline{L}} \\ &= \left(\tfrac{e^{\epsilon}}{(1+\epsilon)^{1+\epsilon}}\right)^{r2^{a-1}/\left((1+\epsilon)2^{a-1}\right)} &&= \left(\tfrac{e^{\epsilon}}{(1+\epsilon)^{1+\epsilon}}\right)^{r/(1+\epsilon)} &&< 0.9421^r.\end{aligned}$$

The event that $L \geq 2^a$ is equivalent to the event that $2^a$ Bernoulli trials are conducted (with expected number of successes $p2^a$) with fewer than $r$ successes. To bound the probability that $L$ is too large, we use (3.2) and the Chernoff bound for deviations below the mean [14, Theorem 4.5, part 1],

$$\begin{aligned}\Pr\left[L \geq 2^{a}\right] &= \Pr\left[r \leq (1-\epsilon')p2^{a}\right] &&\leq \left(\tfrac{e^{-\epsilon'}}{(1-\epsilon')^{1-\epsilon'}}\right)^{p2^{a}} &&= \left(\tfrac{e^{-\epsilon'}}{(1-\epsilon')^{1-\epsilon'}}\right)^{r2^{a}/\overline{L}} \\ &= \left(\tfrac{e^{-\epsilon'}}{(1-\epsilon')^{1-\epsilon'}}\right)^{r2^{a}/\left((1+\epsilon)2^{a-1}\right)} &&= \left(\tfrac{e^{-\epsilon'}}{(1-\epsilon')^{1-\epsilon'}}\right)^{2r/(1+\epsilon)} &&< 0.9421^r.\end{aligned}$$

By the union bound,

$$\Pr[L \notin [2^{a-1}, 2^a)] < 2 \cdot 0.9421^r \tag{3.3}$$

Therefore, by setting $r$ sufficiently large, we can exponentially decrease the probability that $L$ falls outside the range $[2^{a-1}, 2^a)$, independently of $a$. For example, letting $r = 113$ leads to $\Pr\left[L \notin [2^{a-1}, 2^a)\right] < 0.0025$. Since $r$ is a constant depending only on $\delta$, it can be encoded into the tile types as shown in Figure 2.

#### 3.2.2 Computing a Number Exactly using a Sampling Line

As stated previously, our goal is that, with a sampling line of length $O(m^2)$, we can exactly compute a number $m$. The idea is shown in Figure 3, and is inspired by the sampling line of Kao and Schweller [11] but can estimate a number more precisely using a given length, as well as

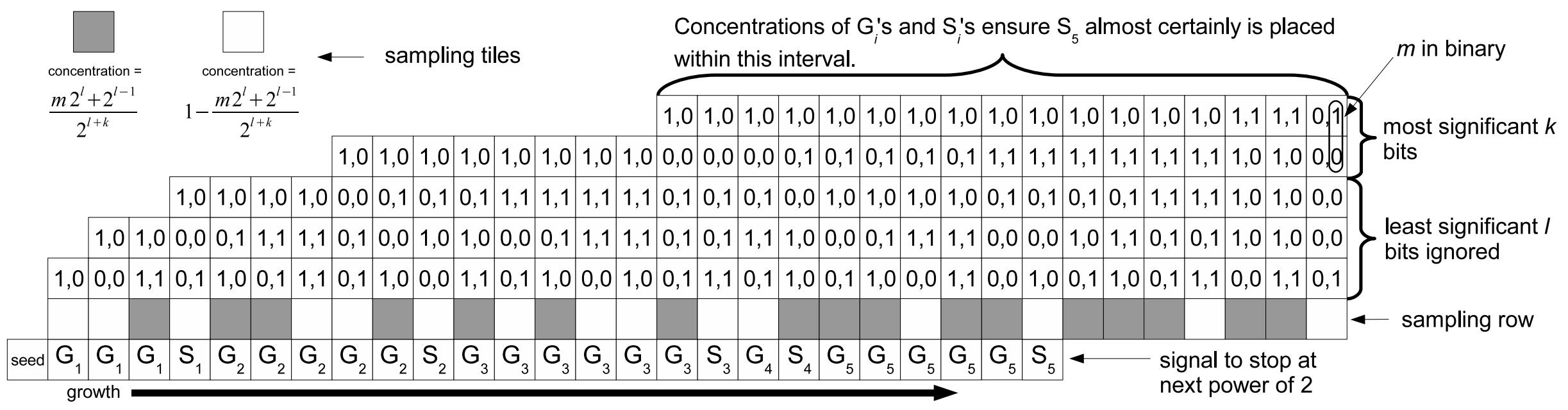

Figure 3: Computing the natural number $m = 2$ from tile concentrations using a sampling line. For brevity, glue strengths and labels are not shown. Each column increments the primary counter, represented by the bits on the left of each tile, and each gray tile increments the sampling counter, represented by the bits on the right of each tile. The number of bits at the end is $l + k$, where $c$ is a constant coded into the tile set, and $k$ depends on $m$, and $l = k + c$. The most significant $k$ bits of the sampling counter encode $m$. In this example, $k = 2$ and $c = 1$.

having a length that is itself controlled more precisely by the technique of Section 3.2.1. The length-controlling portion of the sampling line of length $L$ will control a counter placed above the sampling line, which counts to the next power of 2 greater than $L$, $2^a$. This counter will eventually end up with $a$ total bits before stopping. Let $k$ be the maximum number of bits needed to represent $m$ ($k$ will be about $\frac{1}{3}\log n$ in our application), and let $l = a - k$. We form a row above the row described in Section 3.2.1, which does the sampling. To implement the Bernoulli trials that estimate $m$, one of two tiles $A$ (the gray tile in Figure 3) or $B$ (the white tile in Figure 3) nondeterministically binds to every position of this row. Set the concentration of $A$ to be $\frac{m2^l+2^{l-1}}{2^a}$ and the concentration of $B$ to be $1 - \frac{m2^l+2^{l-1}}{2^a}$. We embed a second counter – the sampling counter – within the primary counter. Whenever $A$ appears, the sampling counter increments, and when $B$ appears it does not change. Let $M$ be the random variable representing the final value of the sampling counter. Then $M$ is a binomial random variable with $\mathrm{E}[M] = m2^l + 2^{l-1}$

We will choose $k$ and $l$ so that the most significant $k$ bits of the sampling counter will almost certainly represent $m$. Intuitively, the least significant $l$ bits of $M$ "absorb" the error. This will occur if $m2^l \leq M < (m+1)2^l$. Note that $m < 2^k$. Let $\varepsilon = \frac{1}{2m}$. Then the Chernoff bound [14, Theorems 4.4/4.5, part 2] and the union bound tell us that

$$\begin{aligned} & \Pr\left[M \geq (m+1)2^l \text{ or } M < m2^l\right] \\ = \ & \Pr\left[M \geq (1+\varepsilon)\mathrm{E}[M] \text{ or } M < (1-\varepsilon)\mathrm{E}[M]\right] \\ \leq \ & e^{-\mathrm{E}[M]\varepsilon^2/3} + e^{-\mathrm{E}[M]\varepsilon^2/2} < e^{-m2^l\left(\frac{1}{2m}\right)^2/3} + e^{-m2^l\left(\frac{1}{2m}\right)^2/2} \\ = \ & e^{-\frac{2^{l-2}}{3m}} + e^{-\frac{2^{l-2}}{2m}} \qquad < e^{-2^{l-k-2}/3} + e^{-2^{l-k-2}/2} \end{aligned}$$

Let $c \in \mathbb{N}$ be a constant. By setting $l = k + c$, the probability of error decreases exponentially in $c$:

$$\Pr\left[M \geq (m+1)2^l \text{ or } M < m2^l\right] < e^{-2^{c-2}/3} + e^{-2^{c-2}/2} < 2 \cdot 0.717^{2^{c-2}}. \tag{3.4}$$

For instance, letting $c = 6$ bounds the left-hand side of (3.4) below 0.0052.

The number of samples is $2^a = 2^{2k+c} = O((2^k)^2)$. Since $m < 2^k$, integers $m$ such that $m^2 \ll n$ can be "probably exactly computed" using much fewer than $n$ Bernoulli trials, and can therefore be computed by a sampling line without exceeding the boundaries of an $n \times n$ square.

### 3.3 Computing $n$ Exactly

We have shown how to compute a number $m$ exactly using a sampling line of length $O(m^2)$ and height $O(\log m)$. To compute $n$, the dimensions of the square, we must compute $m_1, m_2$, and $m_3$, which are the numbers represented by the bits of the binary expansion of $n$ at positions congruent to 1 mod 3, 2 mod 3, and 0 mod 3, respectively. To compute all three of these numbers, we embed two extra sampling counters into the double counter, in addition to the sampling counter described in Section 3.2, to create a quadruple counter. This requires 8 sampling tiles instead of 2, in order to represent each of the possible outcomes of conducting three simultaneous Bernoulli trials, each trial used for estimating one of $m_1$, $m_2$, or $m_3$.

Given $i \in \{1, 2, 3\}$, let $b_i \in \{0, 1\}$ denote the outcome of the $i^{\text{th}}$ of three simultaneous Bernoulli trials, and let $p_i(b_i)$ denote the probability we would like to associate with that outcome. As noted in Section 3.2.2, the values of the $c_i$'s are given by $p_i(1) = \frac{m_i 2^l + 2^{l-1}}{2^a}$, and $p_i(0) = 1 - p_i(1)$.

Since each of the three simultaneous Bernoulli trials is independent, we can calculate the appropriate concentration of the tile representing the three outcomes by multiplying the three outcome probabilities together. Then the required concentration of the tile representing outcomes $b_1, b_2, b_3$ is given by $p_1(b_1) \cdot p_2(b_2) \cdot p_3(b_3)$.

Once the values $m_1$, $m_2$, and $m_3$ are computed, we must remove the $c$ least significant (bottom) bits from the bottom of the primary counter. Since $c$ is a constant depending only on $\delta$, it can be encoded into the tile types. We must then remove the bottom half of the remaining bits.[7] At this point, the concatenation of the bits on the tiles represent the binary expansion of $n$. Rather than expand them out to use three times as many tiles, we simply translate each of them to an octal digit, giving the octal representation of $n$, with one octal digit per tile replacing the three bits per tile. Finally, this representation of $n$ is rotated 90 degrees counter-clockwise, used as the initial value for a decrementing, upwards-growing, base-8 counter, and used to fill in an $n \times n$ square using the standard construction [17]. Rotating $n$ to face up starts the counter $2k+2$ tiles from the bottom of the construction so far. Furthermore, testing whether the counter has counted below 0 requires counting once beyond 0, using 2 more rows than the starting value of the counter. Therefore, to ensure that exactly an $n \times n$ square is formed, the value $n - 2k - 4$, rather than $n$ exactly, is programmed into the tile concentrations to serve as the start value of the upwards-growing counter. An outline of this construction is shown in Figure 4.

### 3.4 Choice of Parameters

We now derive the settings of various parameters required to achieve a desired success probability and derive lower bounds on $n$ necessary to allow the space required by the construction. To ensure probability of failure at most $\delta$, we pick $r$, the number of stages of stopping tiles that must attach before the primary counter is sent the stop signal, so that $2 \cdot 0.9421^r \leq \frac{\delta}{4}$ as in (3.3):

$$r = \left\lceil \frac{\log \frac{\delta}{8}}{\log 0.9421} \right\rceil.$$

[7] Isolating the most significant half of the bits can be done using a tile set similar to the algorithm one might use to program a single-tape Turing machine to compute the function $0^{2n} \mapsto 0^n$.

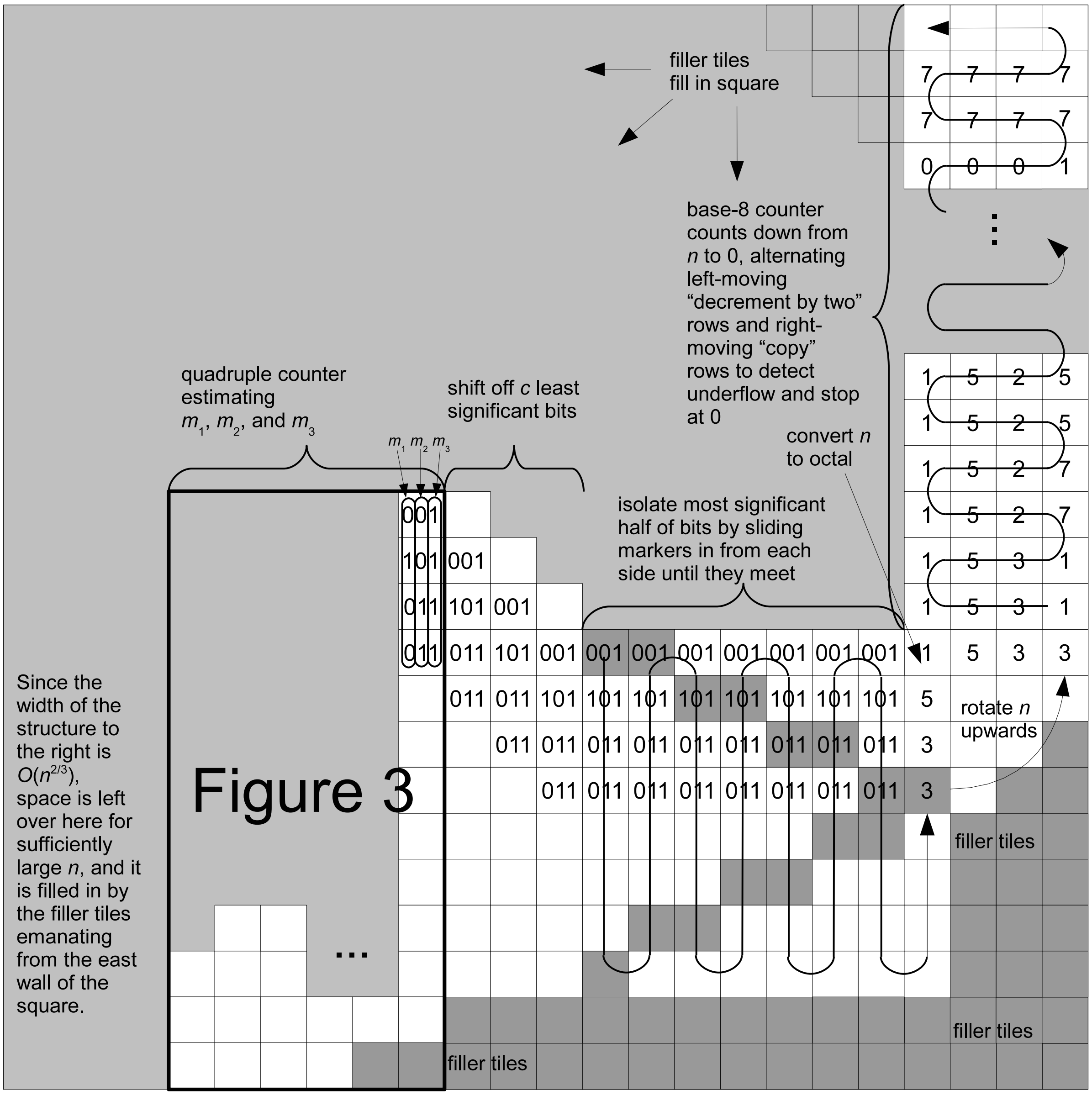

Figure 4: High-level overview of the entire construction, not at all to scale. For brevity, glue strengths and labels are not shown. The double counter number estimator of Figure 3 is embedded with two additional counters to create a quadruple counter estimating $m_1$, $m_2$, and $m_3$, shown as a box labeled as "Figure 3" in the above figure. In this example, $m_1 = 4$, $m_2 = 3$, and $m_3 = 15$, represented vertically in binary in the most significant 4 tiles at the end of the quadruple counter. Concatenating the bits of the tiles results in the string 001101011011, the binary representation of 859, which equals $n - 2k - 4$ for $n = 871$, so this example builds an 871 x 871 square. (Actually, 871 is too small to work with our construction, so the counter will exceed length 871, but we choose small numbers to illustrate the idea more clearly.) Once the counter ends, $c$ tiles ($c = 3$ in this example) are shifted off the bottom, and the top half of the tiles are isolated ($k = 4$ in this example). Each remaining tile represents three bits of $n$, which are converted into octal digits, rotated to face upwards, and then used to initialize a base-8 counter that builds the east wall of the square. Filler tiles cover the remaining area of the square.

For example, choosing $r = 113$ achieves probability of error $\delta/4$ (in ensuring the counter stops between the numbers $2^{a-1}$ and $2^a$) at most 0.0025.

To ensure that each of $m_1$, $m_2$, and $m_3$ are computed exactly, we set $c$, the number of extra bits used in the primary counter beyond $2k$, such that $e^{-2^{c-2}/3} + e^{-2^{c-2}/2} \leq \frac{\delta}{4}$, as in (3.4), or more simply, such that $2 \cdot 0.717^{2^{c-2}} \leq \frac{\delta}{4}$; i.e., set

$$c = 2 + \left\lceil \log \frac{\log \frac{\delta}{8}}{\log 0.717} \right\rceil.$$

For example, choosing $c = 7$ achieves probability of error $\delta/4$ (in ensuring that $m_1$ is computed correctly) at most 0.0025 (in fact, at most 0.000005).

By the union bound, the length of the sampling line and the values of $m_1$, $m_2$, and $m_3$ are computed with sufficient precision to compute the exact value of $n$ with probability at least $1 - \delta$. The example values of $r$ and $c$ given above achieve $\delta \leq 0.01$.

The choices of $r$ and $c$ imply a lower bound on the value of $n$ necessary to allow sufficient space to carry out the construction. Clearly the counter must reach at least value $r$, since there are $r$ different stopping stages. The more influential factor will be the value $c$, which doubles the space necessary to run the counter each time it is incremented by 1. $n$ requires $\lfloor \log n \rfloor + 1$ bits to represent, but our estimation will be a string of length the next highest multiple of 3 above $\lfloor \log n \rfloor + 1$. Therefore, each of $m_1$, $m_2$, and $m_3$ requires

$$k = \left\lceil \frac{\lfloor \log n \rfloor + 1}{3} \right\rceil$$

bits to represent. Recall that the primary counter will have height $2k + c$ and count to $2^{2k+c}$ (so long as $r \leq 2^{2k+c}$). Then, $c$ columns are required to shift off the constant $c$ bits from the least significant bits of the counter, and $2k$ columns are required to shift off the least significant half of the bits of the counter to isolate the $k$ most significant bits. $k$ columns are needed to translate the groups of three bits into octal and to rotate this string to face upwards for the square-building counter.

Hence, the total length required along the bottom of the square to compute $n$ is $\max\{r, 2^{2k+c}\} + c + 3k$. Expanding out the definitions of $r$, $k$, and $c$ derived above gives the lower bound $b(n, \delta)$ on $n$ described in Theorem 3.1.

For sufficiently large $n$ and small enough $\delta$, $r$ is much smaller than $2^{2k+c}$, so the latter term dominates. For example, to achieve probability of error $\delta \leq 0.01$ requires $n > 8{,}000{,}000$. According to preliminary experimental tests, in practice, a smaller value of $c$ is required than the theoretical bounds we have derived. For example, if the desired error probability is $\delta = 0.01$, setting $c = 7$ satisfies the analysis given above, but in experimental simulation, $c = 3$ appears to suffice for probability of error at most 0.01, and reduces the space requirements by a factor of $2^{7-3} = 16$. In this case, $n = 9000$ can be computed by a construction that will stay within the 9000 x 9000 square.

A simulated implementation of this tile assembly system using the ISU TAS Tile Assembly Simulator [15] is available at `http://www.cs.iastate.edu/~lnsa/software.html`. The tile set uses approximately $4500 + 9c + 4r$ tile types, where $r$ and $c$ are calculated from $\delta$ as above.

# 4 Assembly of Finite Scaled Shapes

Soloveichik and Winfree [23] studied the self-assembly of *scaled shapes*, in particular studying the complexity of assembling "scaled-up versions" of finite shapes, as measured by the number of tile types needed to uniquely assemble a scaled shape.

Formally, define a *shape* to be a connected set $S \subseteq \mathbb{Z}^2$. For $c \in \mathbb{Z}^+$, define the *c-scaling* of $S$ to be the set

$$S^c = \left\{ \ (x, y) \in \mathbb{Z}^2 \ \middle| \ (\lfloor x/c \rfloor, \lfloor y/c \rfloor) \in S \ \right\}.$$

Intuitively, $S^c$ is $S$ "magnified by factor $c$". If we imagine that we would like to assemble $S$, but we compromise on assembling $S^c$ instead, then $c$ is referred to as the *resolution loss.* Soloveichik and Winfree proved that for every finite shape $S$, there is a scaling factor $c$ and a TAS $\mathcal{T}$ such that $\mathcal{T}$ uniquely assembles $S^c$, and the number of tile types in $\mathcal{T}$ is within a multiplicative logarithmic factor of the Kolmogorov complexity of $S$, measured as the length in bits of the shortest program outputting a list of the coordinates of $S$.

Kao and Schweller asked whether there is a constant-sized tile set that, through concentration programming, can assemble a scaling of any finite shape with high probability. We answer this question affirmatively, by combining the construction of [23] with the construction of Section 3.2.2.

The following is the main theorem of Section 4.

**Theorem 4.1.** *For all $\delta > 0$, there is a tile assembly system $\mathcal{T}_\delta = (T, \sigma, 2)$ such that, for all finite shapes $S \subset \mathbb{Z}^2$, there exists $c \in \mathbb{Z}^+$ and a tile concentration assignment $\rho_S : T \to [0, \infty)$ such that $S^c$ strictly self-assembles in $\mathcal{T}_\delta(\rho_S)$ with probability at least $1 - \delta$.*

*Proof.* Given a finite shape $S$, the construction of [23] uses an intricate construction of a "seed block" that "unpacks" from the hard-coded tile types a single-tape Turing machine program $\pi \in \{0, 1\}^*$ that outputs a binary string $\text{bin}(S)$ representing a list of the coordinates of $S$ (in fact, a *shortest* program for $\text{bin}(S)$ in the sense of Kolmogorov complexity [13]). The construction of [23] is intended to utilize an asymptotically optimal number of tile types to achieve this unpacking. The width of the seed block is then $c$, chosen to be large enough to do the unpacking, and also large enough to accommodate the simulation of $\pi$ by a tile set that simulates single-tape Turing machines. Once this seed block is in place, a tile set then assembles the scaled shape by carrying $\text{bin}(S)$ through each block, and then using the relative order of the block to determine the next block(s) to assemble. The order in which blocks are assembled is determined by a spanning tree of $S$, so that any blocks with an ancestor relationship have a dependency, in that the ancestor must be (mostly) assembled before the descendant, whereas blocks without an ancestor relationship can potentially assemble in parallel. See [23] for more details.

In a similar fashion to the technique used by Summers [24] to combine the construction of [23] with a temperature programming construction, we replace the seed block tiles of [23] with a tile set that produces the program $\pi$ from tile concentrations, and utilize the remainder of the tile set of [23] unchanged. This is illustrated in Figure 5. For our purpose, we do not require the compactness that necessitated the "unpacking" phase of the construction of [23]. Choose $c$ to be sufficiently large that $\pi$ can be simulated within the trapezoidal region of the $c \times c$ block of Figure 5, and also sufficiently large that the construction of Section 3.2.2 has sufficient room to estimate the binary string $\pi$ from tile concentrations in the center region (the "double counter estimator") of Figure 5. Once this is done, the construction of [23] can take over and assemble the entire scaled shape $S^c$. The portion of the construction of [23] that achieves this is a constant-size tile set, so combined with our construction remains constant. □

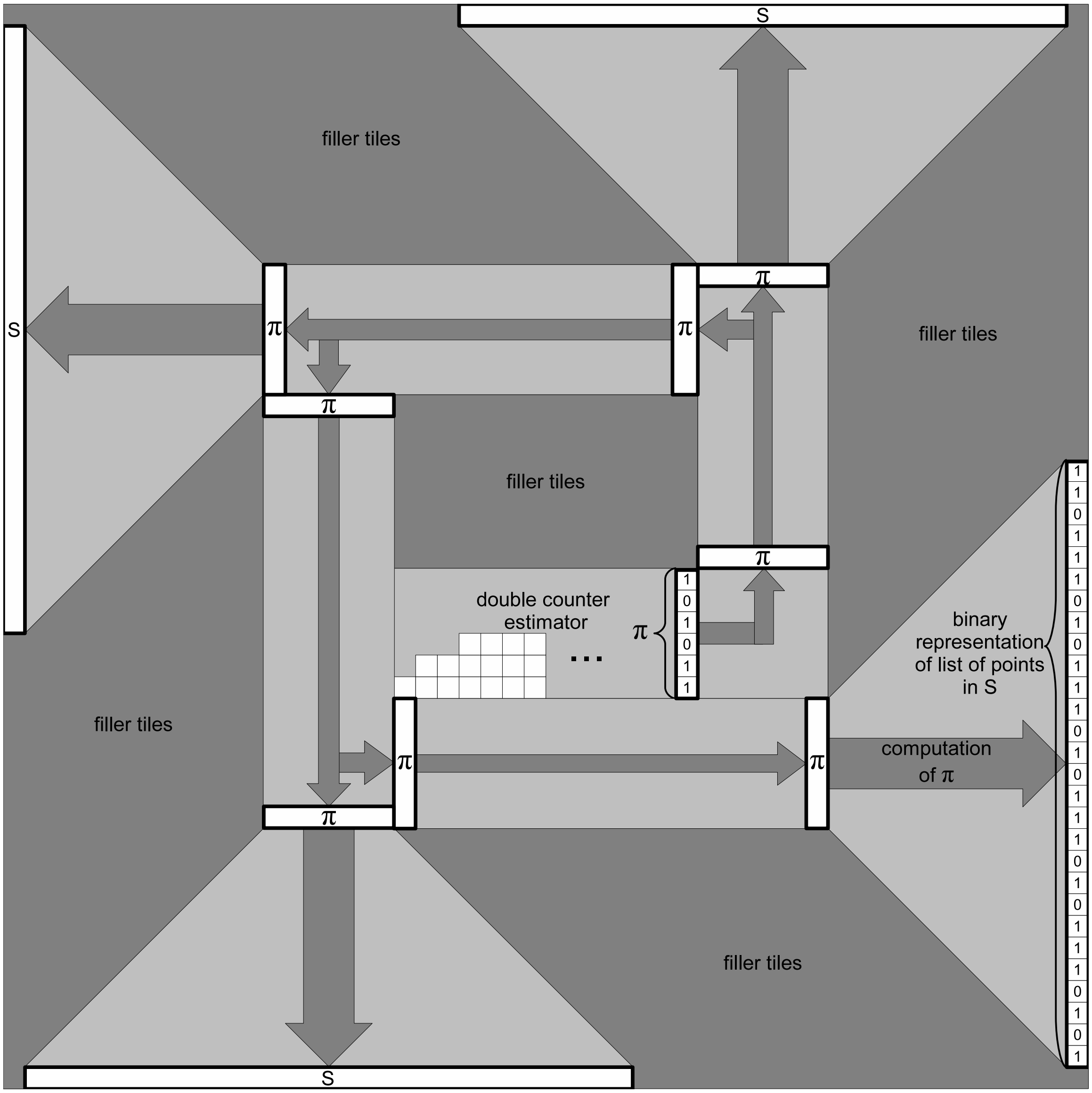

Figure 5: The seed block used to replace the seed block of [23], from which the construction of [23] can assemble a scaled version of the shape $S$ (encoded by a binary string representing the list of coordinates, also labeled "$S$" in the figure), which is output by the single-tape Turing machine program $\pi$. $\pi$ is estimated from tile concentrations as in Figure 3, then four copies of it are propagated to each side of the block, where it is executed in four rotated, but otherwise identical, computation regions. When completed, four copies of the binary representation of $S$ border the seed block, which is sufficient for the construction of [23] to assemble a scaled version of $S$.

The bound on $c$ due to the double-counter estimator construction of Section 3.2.2 is $O\left(2^{|\pi|^2}\right)$. Hence for any shape $S$ whose shortest program takes super-exponential time (in the length of the output, which we may assume is at least $|\pi|$ since $\pi$ is a shortest program for $S$), the resolution loss is no larger than that achieved by [23]. Such shapes are precisely those with greater than exponential *computational depth*, in the sense of Bennett [3].

# 5 Tradeoff between Tile Concentration Precision and Number of Tile Types

## 5.1 Motivation

We have described how a single tile set in Winfree's abstract tile assembly model, appropriately "programmed" by setting tile concentrations, exactly assembles an $n \times n$ square with high probability, for any sufficiently large $n$. This requires specifying tile concentrations to $O(\log n)$ bits of precision (the constant in the $O()$ being about 4 in our construction), which is asymptotically optimal for most $n$ by a standard information-theoretic lower bound.

As observed by Chandran, Gopalkrishnan, and Reif [5], it is perhaps physically unrealistic to enforce that tile concentrations are maintained to an arbitrary degree of precision.[8] They consider a more realistic model of randomized self-assembly in which the system is *equimolar*: all tile concentrations are equal, so that whenever $m \geq 2$ tile types compete to bind to the same position in a growing assembly, each tile type is sampled with uniform probability $\frac{1}{m}$. They show how to construct, for each $n \in \mathbb{Z}^+$, a height-1 line of expected length $n$ using $O(\log n)$ tile types in a randomized equimolar tile assembly system. They show this to be a tight upper bound for *all* $n$ (not just for algorithmically random $n$), and they observe that this is superior to the $n$ tile types required to uniquely assemble a length-$n$ line in the standard deterministic aTAM. In the standard aTAM, Adleman, Cheng, Goel, and Huang [1] similarly specify a number $n$ by assembling an $n \times n$ square, using an optimal number of tile types ($O(\log n / \log\log n)$ in the case of squares).

Intuitively, [1, 5] and the present paper can be thought of as computing (through tile self-assembly) a number $n$ using $O(\log n)$ bits of "input", with the input specified via two extreme approaches:

**present paper:** $O(\log n)$ bits specified optimally in the tile concentrations, and $O(1)$ bits specified in the tile types

**[1, 5]:** $O(1)$ bits specified in the tile concentrations (specifically, 0 bits), and $O(\log n)$ bits specified optimally in the tile types

[8] Though some authors [7, 22] have suggested that it may eventually be possible to control concentrations to the highest precision possible, by controlling *exact molecular counts* of chemicals in solution.

Arbitrary tile concentrations may potentially be too permissive a model, yet it may also be the case that the requirement of equimolar tile concentrations is overly strict. Suppose that a chemist tells us that molecular concentrations can be controlled, but only to $g$ bits of precision for some integer $g$. That is, the chemist can guarantee that if we request a tile concentration of $p \in [0, 1]$, then the actual concentration whenever the tile is sampled will be within $2^{-g}$ of $p$. We must assume that each time the tile is sampled it could potentially be selected with any concentration in the range around $p$; i.e., over the course of assembly the concentration could change according to an adversarial scheme as long as the sampled concentration stays within a distance of $2^{-g}$ of $p$. Then if we wish to estimate a number $n$ requiring $\log n$ bits to describe, we could potentially estimate $g$ of these bits though concentration programming,[9] and the remaining $\log(n) - g$ bits require $\Omega\left(\frac{\log(n)-g}{\log(\log(n)-g)}\right)$ tile types by the Rothemund/Winfree lower bound [17].

### 5.2 Formal Definition of the Model

The construction of Section 3 can be modified to achieve the asymptotic lower bound of Section 5.1. We formalize the conditions stated in Section 5.1 as follows. Let $\epsilon > 0$. Informally, $\epsilon$ represents a *concentration error*, a distance from the desired concentration with which tiles may be sampled. That is, when a tile $t$ is sampled, its concentration could lie anywhere in the range $[\rho(t)-\epsilon, \rho(t)+\epsilon]$. The formal effect of $\epsilon$ on the semantics of assembly are described below.

Modeling multiplicative error in the concentration would mean replacing the above interval with $[\rho(t)(1-\epsilon), \rho(t)(1+\epsilon)]$. Multiplicative error is perhaps a more realistic model than additive error. From the perspective of pure mathematical strength, if $\rho(t) > 1$, then multiplicative error is a stronger constraint, and if $\rho(t) < 1$, then additive error is a stronger constraint. Since we exclusively use concentrations in the interval $(0, 1)$, our results are at least as strong as if we had defined error to be multiplicative. However, multiplicative error gives too much power when $\rho(t) \ll 1$ by allowing us to "cheat" and get around the need for a tradeoff between tile concentrations and tile complexity. In particular, it is still possible to encode an arbitrary number of bits into the tile concentrations, by choosing $\alpha > 1$ and ensuring that for every pair of "potential concentrations" $\rho, \rho'$ (which to use depending on which of two values are being encoded), if $\rho(t) > \rho'(t)$, then $\rho(t)(1-\epsilon) \geq \alpha\rho'(t)(1+\epsilon)$. No matter how large $\epsilon$ is (as long as it is less than 1), $\rho'(t)$ can be set sufficiently small to obey this inequality and allow us to "tell $\rho(t)$ and $\rho'(t)$ apart" even under the error, using a mechanism similar to the construction described in Section 6.3. Additive error is required to impose constraints on tile concentration programming strong enough to truly limit the number of bits that can be encoded into the tile concentrations, so that we are forced to encode some of the bits into the tile types. Since no concentration is above 1, this assumption is not providing us with any extra power lacking under the multiplicative error model. Therefore the results of this section are at least as strong as if we had chosen the multiplicative error model.

To avoid tedious repetition, recall the variables defined in Section 2. Define

$$\rho_\epsilon^-(t) = \max\{0, \rho(t) - \epsilon\}, \qquad \rho_\epsilon^+(t) = \rho(t) + \epsilon,$$

$$f_{\alpha_i,\epsilon}^-(\vec{m}) = \sum_{t \in T_{\alpha_i}(\vec{m})} \rho_\epsilon^-(t), \qquad f_{\alpha_i,\epsilon}^+(\vec{m}) = \sum_{t \in T_{\alpha_i}(\vec{m})} \rho_\epsilon^+(t),$$

[9] Actually, we could potentially estimate some constant multiple of $g$ bits; see the footnote in the proof of Theorem 5.3 for an explanation.

$$p_{\alpha_i,\epsilon}(t|\vec{m}) = \frac{\rho^-_\epsilon(t)}{f^+_{\alpha_i,\epsilon}(\vec{m})}, \quad \text{and} \quad p_{\alpha_i,\epsilon}(\vec{m}) = \frac{f^-_{\alpha_i,\epsilon}(\vec{m})}{\sum\limits_{\vec{n}\in\partial\alpha_i} f^+_{\alpha_i,\epsilon}(\vec{m})}.$$

$\rho$ and $\epsilon$ induce the subprobability measure[10] $P_{\rho,\epsilon} : \mathcal{A}_\Box[\mathcal{T}] \to [0,1]$ defined by

$$P_{\rho,\epsilon}(\alpha) \quad = \sum_{\vec{\alpha}=(\alpha_i|0\le i<k)\in A(\alpha)} \prod_{i=0}^{k-2} p_{\alpha_i,\epsilon}(t_{\vec{\alpha},i}|\vec{m}_{\vec{\alpha},i}) p_{\alpha_i,\epsilon}(\vec{m}_{\vec{\alpha},i})$$

For $p \in [0,1]$, $\epsilon > 0$, and $X \subseteq \mathbb{Z}^2$, we say *$X$ strictly self-assembles in $\mathcal{T}(\rho)$ with probability at least $p$ subject to concentration error $\epsilon$* if $\Pr_{\alpha \xleftarrow{P_{\rho,\epsilon}} \mathcal{A}_\Box[\mathcal{T}]}[\text{dom}\,\alpha \simeq X] \ge p$. That is, $\mathcal{T}$ self-assembles into a shape equal to $X$ with probability at least $p$, even if tile concentrations are maliciously and dynamically adjusted by up to an additive difference of $\epsilon$ throughout the assembly process so as to minimize the probability of assembling $X$.

As discussed in Section 2, we will generally ignore the contribution of $p_{\alpha_i,\epsilon}(\vec{m})$ to the probability of success, since the correctness of the constructions is not affected by choice of frontier location so long as the assembly sequence is fair. Observation 5.1 is an analog of Observation 2.1 using $P_{\rho,\epsilon}$. It strengthens the hypothesis that $\rho(t)$ is strictly positive for all tile types $t$ to the hypothesis that $\rho(t) - \epsilon$ is strictly positive. This ensures that no frontier location $\vec{m}$ is starved due simply to adversarial choice of 0 concentration of the tile types that can bind to $\vec{m}$, implying that our assumption of a fair assembly sequence is justified.

**Observation 5.1.** *Let $\mathcal{T} = (T, \sigma, \tau)$ be a TAS, let $\epsilon > 0$ be a tile concentration error, let $\rho : T \to (\epsilon, \infty)$ be a tile concentration assignment assigning concentrations strictly greater than $\epsilon$, and let $\vec{\alpha}$ be an infinite assembly sequence resulting from the assembly of $\mathcal{T}$ according to $\rho$ and $\epsilon$ as described above. Then* $\Pr[\vec{\alpha}$ is fair$] = 1$.

*Proof.* Similar to the proof of Observation 2.1. □

We term $p_{\alpha_i}(t|\vec{m}) - p_{\alpha_i,\epsilon}(t|\vec{m})$ and $p_{\alpha_i}(\vec{m}) - p_{\alpha_i,\epsilon}(\vec{m})$ the *tile selection probability error* and *frontier selection probability error*, respectively. The model stated above defines error in specifying concentrations, but our proofs require discussing error in tile selection probabilities. The following lemma bounds the latter in terms of the former.

**Lemma 5.2.** *Let $\mathcal{T} = (T, \sigma, \tau)$ be a TAS, let $\epsilon > 0$ be a tile concentration error, let $c = \max\limits_{\alpha\in\mathcal{A}[\mathcal{T}],\vec{m}\in\partial\alpha} |T_\alpha(\vec{m})|$, let $\alpha \in \mathcal{A}[\mathcal{T}]$, let $\vec{m} \in \partial\alpha$ with $f_\alpha(\vec{m}) \ge 1$, let $t \in T_\alpha(\vec{m})$, and let $\rho : T \to [0,\infty)$ be a tile concentration assignment. Then $p_\alpha(t|\vec{m}) - p_{\alpha,\epsilon}(t|\vec{m}) \le (c+1)\epsilon$.*

*Proof.* For any $a, b, \epsilon, \delta \in \mathbb{R}$ such that $\epsilon, \delta > 0$, $b \ge a$, and $b \ge 1$,

$$\begin{aligned} \frac{a-\epsilon}{b+\delta} &= \frac{a}{b} - \frac{a}{b} + \frac{a}{b+\delta} - \frac{\epsilon}{b+\delta} &&= \frac{a}{b} - \frac{a\delta}{b(b+\delta)} - \frac{\epsilon}{b+\delta} \\ \ge \frac{a}{b} - \frac{b\delta}{b(b+\delta)} - \frac{\epsilon}{b+\delta} &= \frac{a}{b} - \frac{\delta+\epsilon}{b+\delta} &&\ge \frac{a}{b} - (\delta+\epsilon). \end{aligned}$$

[10]A *subprobability measure* on a set $X$ is a function $P : X \to [0,1]$ such that $\sum\limits_{x\in X} P(x) \le 1$.

By this inequality with $a = \rho(t)$, $b = f_\alpha(\vec{m})$, and $\delta = c\epsilon$,

$$\begin{aligned} p_{\alpha,\epsilon}(t|\vec{m}) &= \frac{\rho_\epsilon^-(t)}{f_{\alpha,\epsilon}^+(\vec{m})} = \frac{\rho_\epsilon^-(t)}{\sum\limits_{t' \in T_\alpha(\vec{m})} \rho_\epsilon^+(t)} \\ \geq \frac{\rho(t) - \epsilon}{\sum\limits_{t' \in T_\alpha(\vec{m})} (\rho(t') + \epsilon)} &\geq \frac{\rho(t) - \epsilon}{f_\alpha(\vec{m}) + c\epsilon} \geq p_\alpha(t|\vec{m}) - (c+1)\epsilon. \end{aligned}$$

□

In particular, the hypothesis $f_\alpha(\vec{m}) \geq 1$ of Lemma 5.2 holds with equality in our main construction.

## 5.3 Optimal Counter

We recall the construction of an *optimal counter* by Adleman, Cheng, Goel, and Huang [1], which we will combine with the construction of Section 3.3 to prove the theorem. That paper showed how to uniquely assemble an $n \times n$ square from a tile set with $O\left(\frac{\log n}{\log \log n}\right)$ tile types. Much of [1] concerns achieving an optimal running time in addition to an optimal number of tile types, but our current goal is simply to build a square while optimizing the number of tile types and bits of precision of concentration. Therefore we will use a variant of a simpler construction described in [1], rather than their main construction, since we need only to optimize the number of tile types.

Rothemund and Winfree [17] described how to build an $n \times n$ square from $O(1) + \log n$ tile types, by using $\log n$ tile types to encode $n$ in binary, one bit per tile, in a "seed row" that grows immediately from the seed via double bonds.[11] From there a base-2 counter (assembled by a constant number of tile types) counts from $n$ down to 0, and forms one side of the square, from which another constant-sized set of "filler" tiles forms the rest of the square to be as wide as the counter is high, as in Figure 4.

Each of these tile types is an element of a set of cardinality at least $\log n$, yet each is only encoding a single bit, rather than the information-theoretically optimal $\log \log n$ bits. Adleman, Cheng, Goel, and Huang propose choosing an integer $b$ such that

$$b \geq \frac{\log n}{\log \log n},$$

and encoding $n$ in base $b$ in the seed row, using

$$h = \log_b n = \frac{\log n}{\log b} \leq \frac{\log n}{\log \frac{\log n}{\log \log n}} = \frac{\log n}{\log \log n - \log \log \log n} < 2\frac{\log n}{\log \log n}$$

unique tile types. One can then use a base-$b$ counter to imitate the construction of Rothemund and Winfree. However, this counter will no longer use a constant number of tile types, since $b$ depends on $n$. But, by choosing $b$ a power of two such that

$$\frac{\log n}{\log \log n} \leq b < 2\frac{\log n}{\log \log n}, \tag{5.1}$$

[11] The quotes around "seed row" are to indicate that the row is not actually the seed, as the problem is trivialized if one allows a non-constant seed assembly. What we mean by "seed row" is that the tile set is designed so that the first thing to happen during assembly is the attachment via double bonds of tile types hard-coded to bind next to the single seed tile.

we can guarantee that the counter does not use too many tile types either. Specifically, the counter alternates "increment" and "test for overflow" rows as in [17]. For the increment row, for each $d \in \{0, 1, \ldots, b-1\}$, and each $s \in \{\text{MSB}, \text{INTERIOR}, \text{LSB}\}$, and each $c \in \{0, 1\}$, there is a tile type representing "digit $d$ with significance $s$ and borrow value $c$", since no matter the size of $b$, the borrow value is at most 1. For the test for overflow, we require one tile type for each $d \in \{0, 1, \ldots, b-1\}$, and each $s \in \{\text{MSB}, \text{INTERIOR}, \text{LSB}\}$. This uses

$$b \cdot 3 \cdot 2 + b \cdot 3 < 18 \frac{\log n}{\log \log n}$$

tile types. Therefore, combining the seed row tile types and the counter tile types, and using the same constant number of tile types to fill in a square given a counter of the correct height, we require at most $O(1) + 20 \frac{\log n}{\log \log n}$ tile types to assemble an $n \times n$ square.

### 5.4 Tradeoff between Tile Concentration Precision and Number of Tile Types

The following is the main theorem of Section 5.

**Theorem 5.3.** *For all $\delta > 0$, there is a constant $\widehat{c}$ such that the following holds. For all $g, n \in \mathbb{Z}^+$, there is a singly-seeded tile assembly system $\mathcal{T}_{\delta,n,g} = (T, \sigma, 2)$ such that $|T| \leq \widehat{c} + 20 \frac{\log(n) - g}{\log(\log(n) - g)} + \log \log n^{2/3}$, and there is a tile concentration assignment $\rho_n : T \to [0, \infty)$ such that the set $\{\ (x, y) \mid x, y \in \{1, \ldots, n\}\ \}$ strictly self-assembles in $\mathcal{T}_{\delta,n,g}(\rho_n)$ with probability at least $1 - \delta$ subject to concentration error $2^{-g}$.*

That is, $\log(n) - g$, the number of bits remaining in $n$ after $g$ bits have been estimated from concentrations, is asymptotically optimally represented in $T$, by the lower bound of [17].

*Proof.* We combine the construction of Section 5.3 with that of Section 3.3, and the idea is shown in Figure 6. The first non-constant term in the bound on $T$ derives from our modified construction to account for concentration error in the "Bernoulli trial" sampling tiles described in Section 3.2.2. The second non-constant term (negligible compared to the first) comes from fixing the tiles described in Section 3.2.1, and is explained at the end of the proof. First we deal with the Bernoulli trial tiles.

Essentially, the same construction of Figure 4 is used, except that the base-8 counter that formed the east wall of the square is now modified to be a more exotic counter that counts to $n$ using two different bases for the digits, the base used depending on the relative significance of the digit. Intuitively, the least significant digits are in base 8, and are those estimated from concentrations. We choose these digits to be as numerous as possible, while obeying the constraint that they can be precisely estimated even with the concentration error. We call these digits $n_1$ (abusing terminology to refer to the integer and the string of digits as the same object). Suppose $n_1$ is $j/3$ octal digits long (i.e., requires $j$ bits). The remaining portion of $n$ is hard-coded into the tile types, and is used to represent the integer $n_2 = (n - n_1)/2^j$. $n_2$ is represented in base $b$, where $b$ obeys (5.1) with $n_2$ substituted for $n$.

The two strings $n_1$ and $n_2$ are concatenated as in Figure 6, and used as the start value for the unusual counter described earlier. This counter uses two sets of tile types. Those on the right are almost the same as in Figure 4, decrementing in base 8. The difference is that the most significant digit, instead of binding to the "filler" tiles, propagates a borrow to the next column, which is the least significant digit of the optimal counter of Section 5.3. Thus, the optimal counter is decremented once every time the base-8 counter (of width $j/3$) decrements to 0; in other words, the double-base counter counts to the value $n_2 \cdot 8^{j/3} + n_1 = n_2 \cdot 2^j + n_1 = n$.

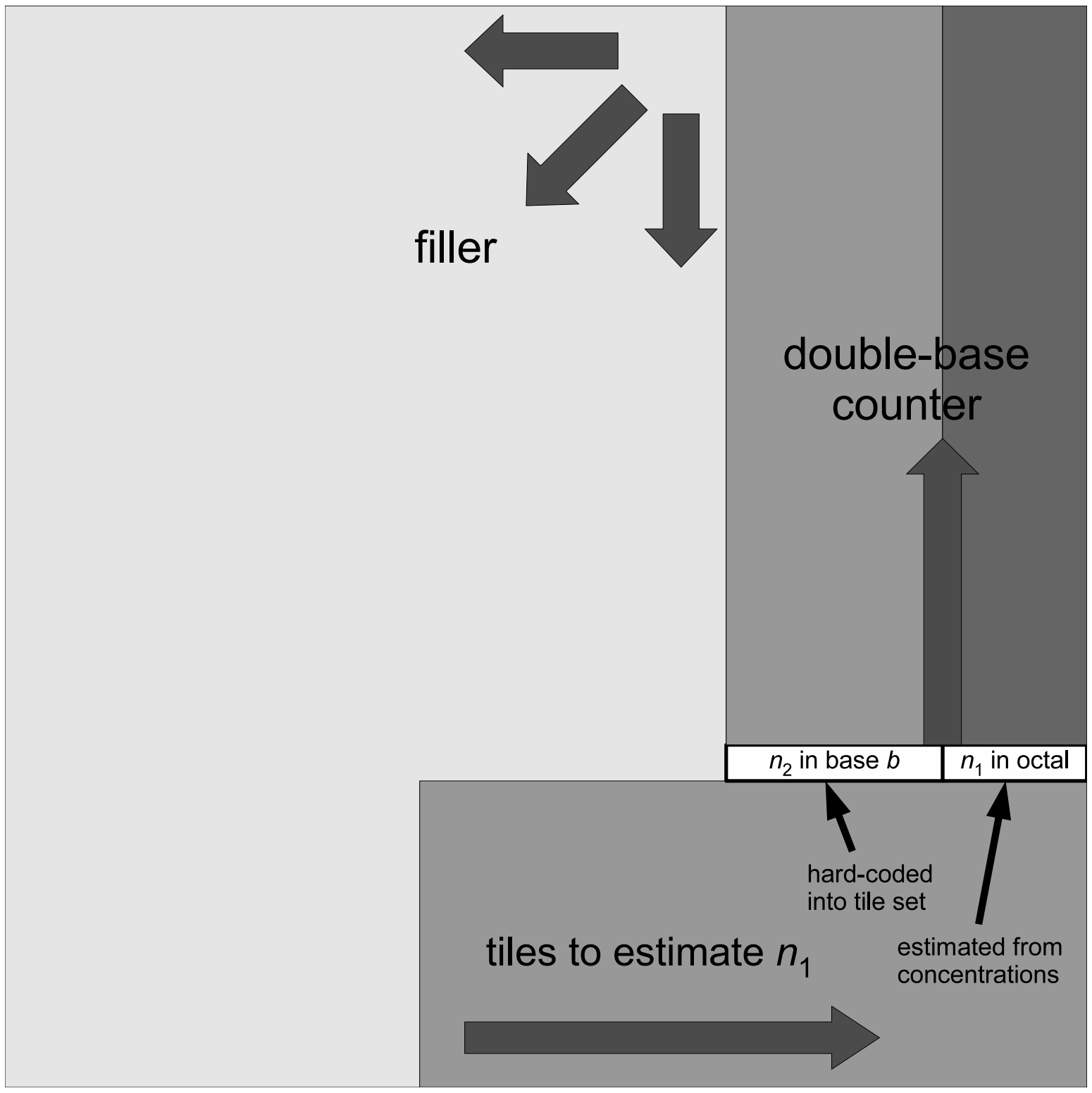

Figure 6: The assembly of an $n \times n$ square taking some bits of $n$ from concentrations, and the rest from the tile types. $n$'s binary expansion is split into $n_1$ and $n_2$. $n_1$ is estimated in base 8 as in Figure 4, and $n_2$ is hard-coded into the tile types in base $b \approx \frac{\log n}{\log \log n}$, as in [1]. These strings are used to drive a "double-base" counter that uses octal digits for the portion representing $n_1$ and base-$b$ digits for the rest, decrementing the base-$b$ portion once for each time the octal counter counts to 0.

It remains to describe how many digits should be allocated between $n_1$ and $n_2$ so as to allow $n_1$ to be estimated with precision even given the concentration error, yet minimize the number of tile types needed to encode $n_2$. In other words, we must choose the maximum value of $j$ (which minimizes the number of tile types for $n_2$) such that a $j$-bit number can be estimated with precision subject to concentration error $2^{-g}$.

First, consider what happens to the construction of Section 3.2.2 if concentration error is introduced. Recall that to estimate a number $m$, the probability of a certain tile $t$ is set to $\frac{m2^l+2^{l-1}}{2^a}$, where $a = O(m^2)$ and $k = a - l$ is the number of bits needed to represent $m$. If $2^a$ trials are conducted, the expected value of $M$, the number of successes, is the midpoint of the interval $[m2^l, (m+1)2^l)$. Consider a concentration error of $\frac{2^{l-6}}{2^a}$. By Lemma 5.2 and the fact that $|T_\alpha(\vec{m})| \leq 8$ for all $\alpha \in \mathcal{A}[\mathcal{T}]$ and all $\vec{m} \in \partial\alpha$, this implies a tile selection probability error of at most $9\frac{2^{l-6}}{2^a} < \frac{2^{l-2}}{2^a}$. In this case, each time $t$ competes to bind, $t$'s probability of being chosen could have expected value as low as $\frac{m2^l+2^{l-2}}{2^a}$ and as high as $\frac{(m+1)2^l-2^{l-2}}{2^a}$, rather than the desired $\frac{m2^l+2^{l-1}}{2^a}$. In other words, the expected value will drift in the middle two quarters of the range $\left[\frac{m2^l}{2^a}, \frac{(m+1)2^l}{2^a}\right)$. A straightforward re-analysis of the Chernoff bounds in Section 3.2.2 shows that in this case, setting $c = l - k$, the error probability can be bounded below $2 \cdot 0.717^{2^{c-4}}$, rather than $2 \cdot 0.717^{2^{c-2}}$ as in Section 3.2.2. That is, the constant in the exponent goes down by 2 to reflect the loss of precision. To make up for this, we must set $c$ to be 2 greater than we would otherwise to ensure that the value $M$ falls in the interval $[m2^l, (m+1)2^l)$ with the same probability.

Considering what this implies for the number of bits that can be reliably estimated from concentrations subject to the error bound, if the concentration error is $2^{-g}$, then we must have $2^{-g} \leq \frac{2^{l-6}}{2^a}$ for the above argument give us the desired error bound. Since $k = a - l$, this is the same as requiring $k \leq g - 6$, where $k$ is the number of bits of $m$. Recall that we combine the estimation of three such integers $m_1$, $m_2$, and $m_3$. Putting their bits together gives us at most $3(g-6)$ bits that can be estimated from concentrations.[12] Let $j = 3(g-6)$.[13]

Let $q = \lfloor \log n \rfloor + 1$ be the number of bits in $n$. The number $n$ is represented in binary as $b_{q-1}b_{q-2}\dots b_0$, where each $b_i \in \{0,1\}$. Let $n_1 = b_{j-1}b_{j-2}\dots b_0$, and let $n_2 = b_{q-1}b_{q-2}\dots b_j$. Then the number of bits of $n_2$ is $q - j$. As argued in Section 5.3, to hard-code these tiles into the tile set so that they bind to the left of $n_1$ as in Figure 6, it suffices to use a number of tile types at most

$$20 \cdot \frac{q-j}{\log{(q-j)}} \quad = \quad 20 \cdot \frac{q-3(g-6)}{\log{(q-3(g-6))}} \leq 20 \cdot \frac{q-g}{\log{(q-g)}},$$

so long as $g \geq 9$. This achieves the first non-constant term of the bound on $|T|$ stated in the theorem.

While we have ensured that the tiles used to conduct Bernoulli trials as in Figure 3 are robust to concentration error, the tiles used in Figure 2 are not so robust. This can be handled in the following way, and is responsible for the second non-constant term of the bound on $|T|$. Observe that the length of the sampling line is a power of two. In our initial construction, it was necessary to program this length entirely into the concentrations to achieve a constant tile set. However, we may now use the fact that this length is succinctly describable to hard-code it into the tile types without increasing the tile complexity by too much. To estimate $n_1$, the sampling line must grow to length $2^a < n_1^{2/3} \leq n^{2/3}$. Hence the length $2^a$ can be described using $\log a < \log\log n^{2/3}$ bits, and therefore as many tile types. Practically, this encoding could be achieved by running a

[12]Technically, for this to be true we would have to split the three sets of Bernoulli trials across three different pairs of competing tile types, rather than combining them using a "product construction" into a single set of eight competing tile types. Such a change to the tile set is easy; for instance, one could run three sampling counters in a row, each one propagating through the bits estimated by the previous sampling counters.

[13]Since we imagine that a concentration error is first fixed, and then we let $n \to \infty$, we could re-phrase part of the statement of the theorem to, "For all $g$, for all sufficiently large $n$," where "sufficiently large" now depends on $g$ as well as the desired error probability $\delta$. Then we could eliminate the need to break up the estimated number into three subsequences, so long as no greater than 1/3 of $n$'s bits are estimated from concentrations. In this case we could choose $j = g - 6$. But we leave the split into three subsequences in the construction, to show that the minimum value of $n$ needed to work can be made not to depend on $g$. However, in our analysis, we drop the coefficient of 3 to make for a simpler theorem statement.

However, as with the idea to improve the upper bound size of the sampling module to $O(n^\epsilon)$ by splitting $n$ into $t$ subsequences, for $t$ a constant, rather than three, we could potentially learn $t \cdot g$ bits from concentrations, rather than just $g$ (even with concentration error fixed to $2^{-g}$), by creating $t$ different pairs of tile types that each sample $g$ bits. The physical basis of this "linear speedup for precision" is that the number of bits learned from concentrations is proportional to the number of tile types created. In other words, the experimenter, to double the number of bits learnable from concentrations, if existing tile types have already been programmed to their maximum precision of concentration, must double the number of tile types in use, and put the same effort into setting the concentrations of those new tile types as precisely as the concentration error allows.

binary counter to count from $a$ (encoded in the tile types using $\log a$ tile types) down to 0, growing upwards, and then using the east wall of this counter, of length $a$, as the first column of a east-growing binary counter that will count to $2^a$, and letting the north wall of this second counter serve the same purpose as the sampling line of Figure 2. ☐

# 6 Unrealistic Aspects of the Concentration Programming Model

This paper solves theoretical problems in a theoretical model and is not intended to be an experimental blueprint. Nonetheless, this section discusses some difficulties with the thermodynamics of physically implementing the concentration programming model, gives arguments that the main construction of this paper is robust to these difficulties, and suggests potential fixes that could help with implementing this construction or other constructions in the concentration programming model.

## 6.1 Concentrations Change as Tiles are Used Up

One obvious observation is that despite the aTAM stipulating an infinite number of copies of each tile type, the number of tiles is finite, and the number will decrease as more and more assemblies are created. If the tiles are not used up at a rate exactly proportional to their concentration, then their concentrations will change.[14]

This potential problem can be overcome by observing that the tile set of Section 3 can be partitioned into "sampling" tiles, which are intended to compete with each other nondeterministically, and "computation" tiles, which are intended to process the results of the sampling. As we argue in Section 5, the sampling tiles are robust to small errors in the actual sampling probabilities. In other words, the concentrations are allowed to vary by a small amount, and the construction will still work. Therefore, we simply set the ratio of sampling tiles to computation tiles to be large, so that by the time all of the computation tiles have been used up (hence stopping any more assemblies from forming), the concentrations of the sampling tiles cannot have changed by very much, even if some were sampled far out of proportion to their concentrations, which is itself unlikely by the law of large numbers.

However, this fix exacerbates the problem discussed in Section 6.2. Fortunately, we argue that these problems may be avoided as well.

## 6.2 Equal Concentrations Required to Approximate the aTAM with the Kinetic Tile Assembly Model

The regular aTAM – even without concentration programming – is not a completely realistic model of what actually happens at the molecular level when DNA tiles interact. It is a useful abstraction, but in reality, chemical reactions are reversible, defying the monotone nature of the aTAM, and DNA sequences binding with "low energy" (i.e., strength less than the temperature) have enough attraction to *partially* overcome the thermal effects that pull tiles apart, defying the constraint that only DNA tiles with sufficiently matched glues will bind.

[14] Note that this difficulty is not unique to tile concentration programming. As explained in Section 6.2, approximating the aTAM in the kTAM requires equal tile concentrations, a requirement that is also challenged by the fact that concentrations may change as tiles are used up during the assembly process.

The kinetic Tile Assembly Model (kTAM), also introduced by Winfree [27], models reality at molecular scales more closely than the aTAM, at the cost of sometimes being more difficult to analyze. A full description is given in [27]; in this paper we only sketch the relevant intuition. The primary differences between the kTAM and aTAM are:

1. Tiles may detach as well as attach.

2. Incorrect tiles may attach.

In fact, the rate at which tiles attach to a frontier location (the *forward rate*) is assumed proportional to their concentration. This would appear to imply that equimolar systems can do no computation whatsoever, as glues no longer affect which tiles go where. The key is this: the rate at which tiles *detach* (the *reverse rate*) is assumed inversely and exponentially proportional to the strength with which they bind. Thus, rather than tiles that bind with strength 2 staying attached forever, and tiles that bind with strength 1 immediately detaching, in the kTAM, tiles with binding strength 2 detach "slowly" and tiles with binding strength 1 detach "quickly". For simplicity, it is assumed that tiles binding with strength 0 detach immediately, so it is as if they never attached. In some papers (for example [6]), the *locking kTAM* model is used, which assumes that tiles bound with strength 3 will never detach, on the assumption that the energy required to break a strength 3 bond is too large to occur due to random thermal fluctuations in any reasonable amount of time. Given these assumptions about strength 0 and $\geq 3$, the remaining difference with the aTAM is that, if an insufficiently attached tile (a tile bound with strength 1) stays on just long enough for another tile to bind and secure it in place with strength 2, then the tile will remain bound for a long time, despite its presence being a potential error.

Winfree showed in [27] that despite the possibility of error, the behavior of any tile assembly system in the aTAM can be approximated arbitrarily closely by the kTAM. The trick is to slow down the assembly process. In the kTAM, there is no explicit temperature parameter, but the relative ratio of the forward rate to the various reverse rates plays a similar role. In particular, if the forward rate is set to be just barely larger than the reverse rate of strength 2 bound tiles, then assembly will drift in a sort of random walk with a small bias towards forward growth, and a large bias toward reverse growth of an insufficiently attached tile. Intuitively, the effect of slowing down the rate of *net* forward growth is to make more time for insufficiently attached tiles to detach before they can become locked in due to a second attachment that secures them with strength 2.

Implicit in this approach is that there is a single well-defined forward rate. But the forward rate is proportional to tile concentrations, so it is not the same for each tile type unless each tile type is equally concentrated. Attempting to vary the relative concentrations will mean either that some tile type will have forward rate much higher than the strength-2 reverse rate, hence will be more likely to get locked in because it makes so many attempts to bind with strength 1 (or will be more likely to lock in some other erroneous attachment), or that some tile type will have forward rate much less than the strength-2 reverse rate, hence will have no net forward growth at all. This argument would appear to doom the utility of the concentration programming model.

Nonetheless, we argue that the construction given in Section 3 – as well of the constructions of [11] and [2], for the same reasons – is robust to this problems. Hence it may be realistic to attempt tile concentration programming with the particular constructions of this paper and [2,11], although the model of tile concentration programming, in general, remains flawed due to the above argument.

The first observation is that all sampling tiles attach with a single strength 2 bond on their input side, and under the standard kTAM model, strength 0 bonds cannot form. Hence there is no possibility of error, in the standard kTAM model, when one of these tiles binds, since there is no other input tile that could be present to, through an insufficient strength-1 attachment, temporarily hold an erroneous tile in place. Of course, given that strength-0 bonds are unrealistic, since even two mismatched glues share some DNA nucleotides in common and will bind with some positive, though small, strength, we must account for even the wrong *kind* of tile being sampled (such as one of the counter tiles attaching to a sampling tile location). The construction has the property that all sampling tiles bind via a single strength-2 bond on their input side, and have no other positive strength glues except at their single output side. Hence, if the wrong kind of tile attaches, it will be held with strength significantly less than 2, no matter what other tiles attach. For instance, if "strength-0" bonds are really "strength-0.1" bonds, then such insufficient attachments will never be held to the existing assembly with greater than strength 1.1. Therefore it is reasonable to claim that such errors will eventually detach, and likely detach quickly, regardless of what other errors collude to increase the likelihood of the erroneous tile remaining. The fix discussed in Section 6.1 should have similar immunity to this problem, as the only errors with enough strength to potentially become locked in are those involving a different sampling tile than the intended binding, but the sampling tiles have concentrations that are relatively close to each other, even if they are much larger than the concentrations of the computation tiles.

### 6.3 Using Concentrations Arbitrarily Close to Uniform

To the extent that the construction still may suffer from the errors caused by variable tile concentrations, there is a fix that can be applied to the construction to alleviate this problem, which will allow the concentrations to be programmed within an arbitarily small interval around 0.5. The construction of Section 3.2.2 uses tiles whose concentrations are set in the following way. First, partition the unit interval into $N$ equal-size subintervals. Set the concentration of a tile type $t$ to be $p$, where $p$ is the midpoint of one of these intervals, and set its rival tile to have concentration $1 - p$. Then count the number of type $t$ tiles that occur in repeated sampling, divide by the number of samples, and determine the subinterval in which that number lies. If enough samples are performed, the sampled probability will almost certainly lie in the same subinterval as $p$.

This entire process could be carried out with probabilities that are arbitrarily close to 0.5, at the cost of requiring extra precision in the programmed concentrations. Simply repeated the entire process described in the previous paragraph, but instead of partitioning the unit interval [0,1], partition the interval $[0.5 - \epsilon, 0.5 + \epsilon]$ for some small $\epsilon > 0$. Using the construction of Figure 3, if $\epsilon = 2^{-d-1}$, then the construction would simply ignore the most significant $d$ bits of the value $M$ except the most significant bit (as well as ignoring the least significant $l$ bits, as in Figure 3). By this trick, the concentrations of the sampling tile types can be made arbitrarily close to uniform.

However, the tile types that determine the length of the sampling line are still given concentrations arbitrarily close to 0 or 1 as $n \to \infty$. In Figure 2, stage $i$ has expected length $\frac{1}{p}$ if the probability of tile type $S_i$ is set to $p \in [0, 1]$. If we want the expected length of a stage to be $L$ this requires setting $p = 1/L$, which approaches 0 for large $L$. This part of the construction may also be adjusted to utilize concentrations arbitarily close to uniform by the following fix, which we describe only informally but is straightforward to implement similarly to the constructions described earlier in this paper.

Let $\epsilon > 0$, and suppose we require that all concentrations must lie in the interval $[0.5 - \epsilon, 0.5 + \epsilon]$.

Choose the smallest positive integer $i$ such that $2^{-i} \leq \epsilon$; we will set concentrations within the interval $[0.5 - 2^{-i}, 0.5 + 2^{-i}]$. Partition the interval $[0.5 - 2^{-i}, 0.5)$ into an infinite number of sub-intervals

$$\begin{aligned} S_i &= [0.5 - 2^{-i}, 0.5 - 2^{-i-1}), \\ S_{i+1} &= [0.5 - 2^{-i-1}, 0.5 - 2^{-i-2}), \\ S_{i+2} &= [0.5 - 2^{-i-2}, 0.5 - 2^{-i-3}), \\ &\ldots \end{aligned}$$

The idea is to set a tile $t$'s concentration $p$ to be in the midpoint of an interval $S_j$ (and set its competing tile $t'$ to have concentration $1-p$), and then, for each $k = i, i+1, i+2, \ldots$, to test whether $t$'s concentration is in the interval $S_k$, stopping at the first $k$ for which this test reports "yes". As we argue later, the very act of conducting these tests will grow a line of appropriate length to conduct the sampling of Section 3.2.2.

If we maintain a count of the total number of appearances of $t$ that appear in a "large" (defined below) power of 2 number of Bernoulli trials, then the frequency with which $t$ appears is within the interval $S_j$ if and only if the most significant $j+1$ bits of the counter that counts the number of appearances of $t$ is the string $01^{j-1}0$. Hence the $k^{\text{th}}$ test is "run a 'large' power of 2 number of Bernoulli trials by letting $t$ and $t'$ compete, count the number of occurrences of $t$, and report 'yes' if the most significant $k+1$ bits of the counter are the string $01^{k-1}0$".

More precisely, recall that the analysis of Section 3.2.2 showed that if we fix a positive integer $k$, conduct $n_k \equiv 2^{k+l_k}$ Bernoulli trials with success probability $p$ set to be in the midpoint of a dyadic interval, where $l_k \equiv k + c$ and $c$ is a constant, then the probability that the $k$ most significant bits of the binary expansion of the number of successes is not equal to the $k$ most significant bits of the binary expansion of $p$ is at most $2 \cdot 0.717^{2^{c-2}}$. We use this fact, but now we define the constant $c$ to depend on $k$ and define it as $c_k \equiv k + c'$, for $c'$ a constant (which will be chosen based on our desired probability $\delta$ of failure of the entire construction). Assuming that the interval $p$ occupies is $S_j$, we want to bound the probability that one of the tests inaccurately identifies $S_k$ as the interval, for $k \neq j$. By the above argument, this probability is at most

$$\begin{aligned} 2\sum_{k=i}^{j} 0.717^{2^{c_k-2}} &< 2\sum_{k=1}^{\infty} 0.717^{2^{c_k-2}} \\ &= 2 \cdot 0.717^{2^{c'-2}} \sum_{k=1}^{\infty} 0.717^{2^k} \\ &< 2 \cdot 0.717^{2^{c'-2}}. \end{aligned}$$

Hence by choosing $c'$ sufficiently large we can make the probability of error sufficiently small to ensure that all tests report the correct answer. It is straightforward to hard-code the constant $c'$ into a tile set as in Figure 4, as is maintaining the number $c_k$ (which begins at the value $c' + i$ and is incremented once for each test). By implementing this tile set to grow below the $x$-axis, the northern boundary of the assembly produced may serve the same function as the northern boundary of the tiles of Figure 2.

It remains to show that the horizontal length to which this structure grows may be controlled to the same precision as the bottom line of Figure 3. Recall that this line could be programmed to grow to a length in the interval $[2^{a-1}, 2^a)$, for arbitrary $a \in \mathbb{Z}^+$. The above construction can be programmed to grow to test an arbitrary number of intervals[15]. By the argument above, the $k^{\text{th}}$ interval test requires $n_k \equiv 2^{k+l_k} = 2^{2k+c_k} = 2^{3k+c'}$ Bernoulli trials. Hence the total number of trials to test intervals $i$ through $j$ is

$$\sum_{k=i}^{j} 2^{3k+c'} = 2^{c'} \sum_{k=i}^{j} 2^{3k}.$$

Since $i$ and $c'$ are constants that depend only on the desired closeness to uniformity of the concentrations and error probability, respectively, we can hard-code small lengths into the tile types and state that for each $j \in \mathbb{Z}^+$, the sampling line can be made to have length

$$\sum_{k=1}^{j} 2^{3k}$$

with high probability. This implies that we may control the length of the sampling line to within a multiplicative factor of 8. By choosing $d = 2^{3k}$ a power of 8 such that $2^a \leq d \leq 8 \cdot 2^a$ for the value $a$ in Section 3.2.2, the sampling line will have sufficient length to achieve the error probability bounds of Section 3.2.2, yet will still have length bounded by $O(n^{2/3})$, hence will fit inside the $n \times n$ square.

Thus all concentration-programmed tile types, both the tile types of Section 3.2.2 doing the Bernoulli trial sampling and the tile types of Section 3.2.1 determining the number of samples, can be assigned concentrations arbitrarily close to uniform, hence allowing an arbitrarily close approximation of Winfree's approximation.

# 7 Conclusion

## 7.1 Potential Improvements

The focus of the present paper is on conceptual clarity. We have therefore described the simplest (i.e., easiest to understand, but not necessarily smallest) version of the tile assembly system that achieves the desired *asymptotic* result that an $n \times n$ square assembles with high probability for sufficiently large $n$. We now observe that this theoretical result could be improved in practice by complicating the tile set.

Our implementation of the tile set uses approximately $4500 + 9c + 4r$ tile types, where, for example, $r = 113$ and $c = 7$ are sufficient to achieve error probability $\delta \leq 0.01$. The tiles are so numerous because of the need to simultaneously represent 4 bits in a tile, in addition to information such as the significance of the bit (MSB, LSB, or interior bit), and doing computation such as addition, which requires tiles that can handle the $2^8$ possible input bit + carry signals. Putting together a few such modules of tile sets results in thousands of tiles before too long. The number of tile types could be reduced by splitting the estimation of $m_1$, $m_2$, and $m_3$ into three distinct geometrical regions, so that each tile is required to remember less information. This would complicate the tile set, as it would require more shifting tricks to ensure sufficient room for all counters, and would require bringing the bits back together again at the end, but it would likely reduce the number of tile types.

[15]Specifically, if we wish to quit after testing precisely $b$ intervals, set $p$ to lie in the midpoint of the interval $S_{i+b-1}$

A large value of $n$ is required to achieve a probability of success at least $(1-\delta)$ for reasonably small $\delta$; $n > 8 \cdot 10^6$ is required to estimate $n$ with 99% chance of correctness (in theory, but apparently not in practice, as discussed in Section 3.4). This shortcoming can be compensated in a number of ways.

In a similar spirit to the linear speedup theorem, more than three simultaneous Bernoulli trials may be conducted with each sampling tile. For example, conducting 6 Bernoulli trials with each sampling tile would estimate two bits of $m_1, \ldots, m_3$ with per sampling tile, rather than one bit, halving the required length of the sampling line. This would result in a prohibitively large tile set, however; as the number of tile types increases exponentially with the number of simultaneous Bernoulli trials per tile type.

A conceptually simpler and practically more feasible improvement is to use 0/1-valued tile concentrations to simulate tile *type* programming (i.e., designing tile types specially to build a particular size square, as in [17]) for small values of $n$, by including tile types that deterministically construct an $n \times n$ square for each small $n$, setting concentrations of those tiles to be 1 and setting concentrations of all other tiles to be 0. Many of the same square-building tile types for can be reused for different values of $n$ (see [17]), with the different values of $n$ largely being dependent on $\approx \log n$ hard-coded tiles that immediately attach to the seed. For singly-seeded tile systems, $3 \log N + O(1)$ tiles are required to handle all $n \leq N$: for each $i \in \{1, \ldots, \log N\}$ that represents the position of a bit of $n$, three tile types are required, one representing "0 at position $i$", one representing "1 at position $i$" (each of which has double-strength bonds on two sides), and one representing "end of string at position $i$" (with a double-strength bond on one side and a zero-strength bond on the other). Though this solution lacks the "feel" of tile concentration programming, it is likely that real-life implementations of tile concentration programming will need to use such hard-coding tricks for smaller structures that lack the space to carry out the amount of sampling required to reconstruct precise inputs solely from tile concentrations.

An alternate improvement to the tile set would be to combine the present technique with the Kao-Schweller technique of building a sampling line inside of a square, to more efficiently use the $n^2$ space available to carry out the estimation. However, square-building is not necessarily the only application of this technique, as shown in Section 4.

The primary novel contribution of this paper is a tile set that, through appropriate tile concentration programming, forms a thin structure of length $O(n^{2/3})$ and height $O(\log n)$,[16] whose rightmost tiles encode the value of $n$ in binary. This binary string could be used to assemble useful structures other than squares, such as rectangles and other supersets of the sampling structure that are "easily encoded" in a binary string of length $O(\log n)$. For the task of building a square, this construction wastes the $\approx n^2$ space available above the thin rectangle, but for computing other structures, it may be advantageous that the rectangle is kept thin. For instance, biochemists routinely use filters (e.g., Millipore Ultrafiltration Membranes) and porous resins [4] to separate proteins based on size, in order to isolate one particular protein for study. The ability to precisely control the size of the filter holes or resin beads would allow for more targeted filtering of proteins than is possible at the present time. DNA is likely too reactive with amino acids to be used as the substrate for such a structure, so an implementation of the tile assembly model not based on DNA

[16] By partitioning $n$'s binary representation into $t$ rather than three subsequences, for $t \in \mathbb{N}$ a constant, the number of trials needed to estimate $n$ is $O(n^{2/t})$. However, the constant factors in the $O()$ increase, making the technique even less feasible for small values of $n$. But if some application requires an asymptotically very short line, the line can be made length $O(n^{\varepsilon})$ for any $\varepsilon > 0$ using this technique.

would be required for such a technique.

Similarly, polyacrylamide gel electrophoresis [21], another technique for discriminating biological molecules on the basis of size, requires molecular mass size markers, which are control molecules of known molecular mass, in order to compare against the molecule of interest on the gel. At the present time, some naturally-occurring molecules of known mass are used, but their masses are not controllable, and the ability to quickly and easily assemble molecules of precisely a desired target mass would be useful in experiments requiring mass markers that differ from the standards. Again, DNA is a special case in which this idea is unnecessary, since precise standards have been developed for DNA gels (e.g., Novagen DNA Markers). But the tile assembly model may one day be implemented using substances that are appropriate for a protein gel.

### 7.2 Open Questions

The proof of Theorem 3.1 shows that for every $\delta, \epsilon > 0$, a tile set exists such that, for every $n \in \mathbb{N}$, appropriately programming the tile concentrations results in the self-assembly of a structure of size $O(n^\epsilon) \times O(\log n)$ whose rightmost tiles represent the value $n$ with probability at least $1 - \delta$. Is this optimal?

Formally, say that a tile assembly system $\mathcal{T} = (T, \sigma, 2)$ is *$\delta$-concentration programmable* (for $\delta > 0$) if there is a (total) computable function $r : \mathcal{A}_\Box[\mathcal{T}] \to \mathbb{N}$ (the *representation function*) such that, for each $n \in \mathbb{N}$, there is a tile concentration assignment $\rho : T \to [0, \infty)$ such that $\Pr[r(\mathcal{T}(\rho)) = n] \geq 1 - \delta$. In other words, $\mathcal{T}$, programmed with concentrations $\rho$, almost certainly self-assembles a structure that "represents" $n$, according to the representation function $r$, and such a $\rho$ can be found to create a high-probability representation of *any* natural number. In the case of the construction of Section 3, $r(\mathcal{T}(\rho))$ outputs the integer represented in binary on the right side of the structure of Figure 3 (and fits into space $O(n^{2/3})$ if the tile types described in Figure 4 are not included in $T$).

**Question 7.1.** *Is the following statement true? For each $\delta > 0$, there is a tile assembly system $\mathcal{T}$ and a representation function $r : \mathcal{A}_\Box[\mathcal{T}] \to \mathbb{N}$ such that $\mathcal{T}$ is $\delta$-concentration programmable and, for each $\epsilon > 0$ and all but finitely many $n \in \mathbb{N}$, $\Pr[|\text{dom } \mathcal{T}(\rho)| < n^\epsilon] \geq 1 - \delta$. If so, what is the smallest bound that can be written in place of $n^\epsilon$?*

In other words, we are asking if the $O(n^\epsilon)$ upper bound on the size of the self-assembled structure representing $n$ that is obtained in the proof of Theorem 3.1 is optimal. That structure has size $O(n^{2/3})$, and for each $\epsilon > 0$, the construction could be modified to have size $O(n^\epsilon)$. Is there a *single* construction whose size is at most $n^\epsilon$ for all $\epsilon > 0$, for sufficiently large $n$? $\Omega(\log n)$ is a clear lower bound on the size of the structure, as it requires $\log n$ space to represent most integers $n$, but it would be interesting to find a larger lower bound than $\Omega(\log n)$, or a smaller upper bound than $O(n^\epsilon)$.

The next question is less formal. Section 6 discusses unrealistic aspects of the concentration programming model, and goes into the detail of the construction of Section 3 to explain why that particular construction does not suffer from the problems associated with these unrealistic assumptions. However, a good model of reality is one that requires no excessively unrealistic assumptions, in which conclusions reached within the model can be inferred to apply to reality without having to inspect the detail of the argument leading to the conclusion.

**Question 7.2.** *Is there a model of concentration programming that "automatically avoids" the*

*problems discussed in Section 6, but which retains (some of) the power of the constructions of [2], [11], and the present paper?*

A better model of concentration programming would free future tile concentration programming constructions from requiring the equivalent of Section 6 of the current paper.

**Acknowledgments.** I thank Jack Lutz for helpful advice during the preparation of this paper, and Julie Hoy for suggesting potential biochemical applications. I appreciate Satyadev Nandakumar, Robbie Schweller, Serge Randriambololona, Walid Mnif, and Ehsan Chiniforooshan, for valuable discussions about several parts of the paper. I especially appreciate the comments of anonymous referees and Ehsan Chiniforooshan for insightful discussions helping to resolve the issues raised by the referees. I also thank Pavan Aduri and Srikanta Tirthapura for allowing me to audit their classes on randomness and computation, whose subject matter proved very useful in devising the techniques of this paper. I am indebted to Erik Winfree for pointing out the problems mentioned in Section 6 and to both Erik Winfree and David Soloveichik discussing them with me. This research was supported in part by National Science Foundation (NSF) Grants 0652569 and 0728806 to Jack Lutz, by NSF grant CCF:0430807 to Pavan Aduri, by Natural Sciences and Engineering Research Council of Canada (NSERC) Discovery Grant R2824A01 and the Canada Research Chair Award in Biocomputing to Lila Kari, and by the Computing Innovation Fellowship grant to David Doty.